\documentclass{article}
\pdfoutput=1
\usepackage[a4paper, total={6in, 8in}]{geometry}
\usepackage[onehalfspacing]{setspace}
\usepackage[utf8]{inputenc}
\usepackage{amsmath}
\usepackage{graphicx,psfrag,epsf}
\usepackage{enumerate}
\usepackage{natbib}
\usepackage{authblk}
\usepackage{url} 
\usepackage{xcolor}
\usepackage{float}
\newcommand{\bm}{\boldsymbol}

\DeclareMathOperator{\diag}{diag}

\DeclareMathOperator*{\argmin}{{arg\,min}}

\def \wvec {\text{\boldmath$w$}}    
\def \xvec {\mathbf{x}}    
\def \yvec {\mathbf{y}}    
\def \zvec {\mathbf{z}}

\def \alphavec        {\text{\boldmath$\alpha$}}
\def \betavec         {\text{\boldmath$\beta$}}
\def \gammavec        {\text{\boldmath$\gamma$}}

\def \epsilonvec      {\text{\boldmath$\epsilon$}}
\def \varepsilonvec   {\text{\boldmath$\varepsilon$}}

\def \etavec          {\text{\boldmath$\eta$}}
\def \thetavec        {\text{\boldmath$\theta$}}

\def \tauvec          {\text{\boldmath$\tau$}}

\usepackage{algorithm}
\usepackage{amssymb}
\usepackage[]{algpseudocode}
\usepackage{caption,subcaption,tabularx,booktabs}
\usepackage[]{threeparttable}

\title{Efficient variational approximations for state space models\thanks{
We would like to thank Wei Wei, Klaus Ackermann  and Ashley Andrews for helpful discussions. Rub\'en Loaiza-Maya gratefully acknowledges
support by the Australian Research Council through grant DE230100029.} }
\date{\today}
\author{Rub\'en Loaiza-Maya\thanks{
Correspondence to: Department of Econometrics \& Business Statistics, Monash University, Clayton VIC 3800, Australia, e-mail: \textsf{ruben.loaizamaya@monash.edu}} }
\author{Didier Nibbering}
\affil{\small Department of Econometrics and Business Statistics, Monash University}

\begin{document}
\maketitle

\begin{abstract} 
\noindent 
Variational Bayes methods are a potential scalable estimation approach for state space models. However, existing methods are inaccurate or computationally infeasible for many state space models. This paper proposes a variational approximation that is accurate and fast for any model with a closed-form measurement density function and a state transition distribution within the exponential family of distributions. We show that our method can accurately and quickly estimate a multivariate Skellam stochastic volatility model with high-frequency tick-by-tick discrete price changes of four stocks, and a time-varying parameter vector autoregression with a stochastic volatility model using eight macroeconomic variables.


\end{abstract}
{\bf Keywords}: State space models, Variational Bayes, Stochastic volatility, Multivariate Skellam model, Time-varying parameter vector autoregression
\\
{\bf JEL Classification:} C11, C22, C32, C58

\thispagestyle{empty}
\clearpage
\setcounter{page}{1}

\section{Introduction}
Estimation of many state space models with nonlinear and/or non-Gaussian measurement equations is computationally challenging \citep{gribisch2022modeling,chan2022large,cross2021macroeconomic}. The likelihood function of these models involves a high-dimensional integral with respect to the state variables which cannot be solved analytically, and hence renders maximum likelihood estimation to be infeasible. As an alternative, exact Bayesian estimation methods allow for the computation of the posterior distribution of the model parameters. These methods either use particle filtering \citep{chopin2020introduction}, or sample from the augmented posterior of the model parameters and the states 
using analytical filtering \citep{carter1994gibbs}. Both approaches can become computationally costly, especially with high-dimensional state vectors or when dependence between the states and the parameters is strong \citep{quiroz2018gaussian}.

Variational Bayes (VB) methods can provide a scalable alternative to exact Bayesian methods. Instead of sampling exactly from the posterior, VB calibrates an approximation to the posterior via the minimization of a divergence function. However, off-the-shelf variational methods for state space models, such as mean-field variational approximations, are known to be poor \citep{wang2004lack}. Gaussian VB methods as proposed by \citet{tan2018gaussian} and \citet{quiroz2018gaussian} use a variational family for the states that conditions on the model parameters and not on the data. The inaccuracy of these existing methods is due to the quality of the approximation to the conditional posterior distribution of the states \citep{frazier2022variational}. More accurate VB methods are computationally infeasible for many state space models. For instance, \citet{tran2017variational} exactly integrate out the states in the variational approximation using particle filtering. The method of \citet{loaiza2022fast} is designed for the specific class of state space models where generation from the conditional posterior of the states is computationally feasible.


This paper proposes a novel VB method that is accurate and fast, and can be applied to a wide range of state space models for which estimation with existing methods is either inaccurate or computationally infeasible. Our method uses a variational approximation to the states that directly conditions on the observed data, and as such produces an accurate approximation to the exact posterior distribution. The approach is faster than existing VB methods for state space models due to the computationally efficient calibration steps it entails. The implementation only requires a measurement equation with a closed-form density representation, and a state transition distribution that belongs to the class of exponential distributions. This allows for a wide range of state space models, including ones with nonlinear measurement equations, certain types of nonlinear transition equations, and high-dimensional state vectors.

Our approximation to the states is the importance density proposed by \citet{richard2007efficient} in the context of efficient importance sampling. Hence, we refer to our method as Efficient VB. \citet{scharth2016particle} employ this importance distribution within a particle Markov chain Monte Carlo (PMCMC) sampler to reduce the variance of the estimate of the likelihood function. The use of this importance density inside PMCMC does not result in substantial computational gains, as it must be recalibrated at each iteration. 
Since VB poses an optimization problem, we can use stochastic gradient ascent (SGA) instead of a sampling algorithm. Our SGA algorithm requires draws from the approximation to the states, which is used to construct an estimate of the gradient of the objective function with respect to the parameters.
Since the importance density is easy to generate from, and it does not have to be recalibrated at each SGA step, the optimization routine is fast and hence scalable to state space models with high-dimensional state vectors and a large number of observations.

Numerical experiments show that the proposed Efficient VB method provides accurate posterior densities, while it only takes a fraction of the computational cost of MCMC.
The experiments employ the stochastic volatility model as the true data generating process. Since the exact posterior can be computed by MCMC methods, the accuracy of our method can be assessed for this model. 
We find that Efficient VB produces variational approximations to the states that are close to the exact posteriors, which result in accurate variational approximations to the parameters of the model. Efficient VB is faster than all benchmark methods with all sample sizes under consideration.

To illustrate the contributions of our method, we apply it in two empirical applications. The first application fits a multivariate Skellam stochastic volatility model with high-frequency tick-by-tick discrete price changes of four stocks. 
With the recent availability of high frequency trading data, the modelling of tick-by-tick  price changes has become increasingly popular \citep{shephard2017continuous,koopman2018dynamic,catania2022dynamic}.
The model we are estimating is in the spirit of the univariate Skellam stochastic volatility model of \citet{koopman2017intraday} but extended to the multivariate setting.
This state space model with a non-linear measurement equation and a multivariate state vector cannot be accurately estimated with existing methods in a reasonable amount of time. Efficient VB produces posterior distributions close to the exact posteriors. 


The second empirical application fits a state space model with Efficient VB that can also be estimated with existing, computationally costly, VB methods.  
This application fits a time-varying parameter vector autoregression with a stochastic volatility model to eight macroeconomic variables. This high-dimensional time series model is proposed by \citet{huber2021inducing}, and related models are used by for instance \citet{clark2015macroeconomic} and \citet{carriero2019large}. This is a state space model with a nonlinear measurement equation and a high-dimensional state vector. In this complex model, our approach is accurate 
while the computation time is a fraction of the computation time of the benchmark methods. 

The proposed VB method has the potential to produce fast and accurate estimation for a wide range of models. Computationally challenging state space models are currently estimated by VB methods that are limited to specific state space formulations. For instance, 
\cite{chan2022fast} and \cite{gefang2019variational} propose a VB method for a specific class of vector autoregression models. \cite{koop2018variational} propose a VB method for a class of time-varying parameter models. 

Existing variational inference methods for state space models that construct point estimates for model parameters, instead of posterior distributions, are computationally expensive.
For instance, \cite{naesseth2018variational} construct an approximation to the conditional posterior of the state using particle filtering, which is computationally costly and hence hinders the scalability to problems with high-dimensional state vectors. \cite{archer2015black} use neural networks to construct an approximation to the posterior of the states. The parameters of these neural networks are calibrated jointly with the parameters of the model, which means that a high-dimensional gradient has to be computed in each iteration of the optimization algorithm.

The outline of the remainder of this paper is as follows. Section~\ref{sec:ssm} discusses specification and exact estimation of state space models, and Section~\ref{sec:vb} develops our VB method. Section~\ref{sec:simulation} conducts numerical experiments to evaluate its accuracy and computational costs, and Section~\ref{sec:skellam} and \ref{sec:application} apply our method to real data. Section~\ref{sec:conclusion} concludes.

\section{State space models}\label{sec:ssm}
Let $\mathbf{y}=(\mathbf{y}_1^\top,\dots,\mathbf{y}_T^\top)^\top$ be an observed time series assumed to have been generated by a state space model with measurement and state densities
\begin{align}
    \mathbf{y}_t|(\mathbf{X}_t=\mathbf{x}_t) &\sim p(\mathbf{y}_t|\mathbf{x}_t,\bm\theta), \label{eq:measurement}\\
    \mathbf{X}_t|(\mathbf{X}_{t-1}=\mathbf{x}_{t-1}) &\sim p(\mathbf{x}_t|\mathbf{x}_{t-1},\bm\theta), \label{eq:state}
\end{align}
respectively, and where the prior density for $\mathbf{X}_1$ is $p(\mathbf{x}_1|\bm\theta)$, $\bm{\theta }\in\Theta$ is a $d$-dimensional parameter vector and $\mathbf{y}_t$ is an $N$-dimensional observation vector with $t=1,\dots,T$. The likelihood function for this  model is given by
\begin{align}\label{eq:likelihood}
p(\mathbf{y}|\bm\theta)=\int p(\mathbf{y},\mathbf{x}|\bm\theta)d\mathbf{x},
\end{align}
where $\mathbf{x} = (\mathbf{x}_1^\top,\dots,\mathbf{x}_T^\top)^\top$ and $p(\mathbf{y},\mathbf{x}|\bm\theta) = \prod_{t=1}^{T}p(\mathbf{y}_{t}|\mathbf{x}_{t},{\bm\theta })p(\mathbf{x}_{t}|\mathbf{x}_{t-1},{\bm\theta })$. Typically, the integral that characterises the likelihood is intractable, as it does not have an analytical solution. This is the case for state space models that assume non-linear or non-Gaussian measurement equations. These types of models are pervasive in econometrics and include, for instance, stochastic volatility models, some time-varying parameter models, and states space models for discrete data. Hence, maximum likelihood estimation is infeasible for a large class of econometric problems. 

Bayesian analysis is concerned with computing the posterior density $p(\bm{\theta }|\mathbf{y})\propto p(\mathbf{y}|\bm\theta)p(\bm{\theta })$, where $p(\bm{\theta })$ is a given choice of prior density. The intractability in the likelihood function is tackled via two different avenues.  First, for certain state space models it is feasible to use Markov chain Monte Carlo (MCMC) to generate from the augmented density
\begin{align}\label{eq:augmented_posterior}
    p(\bm\theta,\mathbf{x}|\mathbf{y}) \propto p(\mathbf{y},\mathbf{x}|\bm\theta)p(\bm\theta),
\end{align}
where analytical filtering methods are used to obtain draws from $p(\mathbf{x}|\bm\theta,\mathbf{y})$. MCMC effectively samples from $p(\bm\theta|\mathbf{y})$ which is a marginal density of $p(\bm\theta,\mathbf{x}|\mathbf{y})$. This approach is limited to certain classes of state space models, and the filtering techniques used can become computationally costly for large sample sizes or high-dimensional state vectors. The second Bayesian avenue generates samples from the posterior by replacing the likelihood function by its unbiased estimate  $\widehat{p}_S(\mathbf{y}|\bm{\theta})$.
This unbiased estimate, evaluated via particle methods, is then used inside a Metropolis-Hastings scheme. This approach, known as particle MCMC, trades off accuracy in estimation of $p(\bm\theta|\mathbf{y})$ by computational speed, via the choice in the number of particles $S$ \citep{andrieu2010particle,doucet2015efficient}. While it can be applied to a broad class of state space models, this approach is known to be computationally costly and highly noisy for an inadequately low number of particles. This issue is exacerbated when the state vector is high-dimensional and a larger number of particles is required.

\section{Variational Bayes}\label{sec:vb}
Variational Bayes may overcome the computational challenges in estimating state space models. The general idea behind VB is to approximate the exact posterior $p(\bm{\theta}|\mathbf{y})$ with an approximating density $q_{\hat{\lambda}}(\bm\theta)\in\mathcal{Q}$, where $\mathcal{Q} = \{q_{\lambda}(\bm\theta): \bm\lambda\in\Lambda\}$ is a class of tractable approximating densities indexed by the variational parameter $\bm{\lambda}\in\Lambda$. The most popular choice for $\mathcal{Q}$ is the Gaussian distribution class. The optimal variational parameter $\hat{\bm{\lambda}}$ is then calibrated by finding the element in $\mathcal{Q}$ that minimizes the Kullback-Leibler (KL) divergence - or any other divergence - to the exact posterior. Implementation of VB requires evaluation of the likelihood function $p(\mathbf{y}|\bm{\theta})$, which is infeasible for most state space models. \cite{tran2017variational} circumvent this issue by replacing $p(\mathbf{y}|\bm{\theta})$ by the unbiased estimate $\widehat{p}_S(\mathbf{y}|\bm{\theta})$. While this approach is faster than PMCMC, it remains computationally costly due to its use of particle filtering.

VB can circumvent the computational challenges of exactly integrating out $\mathbf{x}$, by instead constructing an approximation to the augmented posterior in \eqref{eq:augmented_posterior}. In this case, the approximating density is $q_{\hat{\lambda}}(\bm\theta,\mathbf{x})$ and $\mathcal{Q} = \{q_{\lambda}(\bm\theta,\mathbf{x}): \bm\lambda\in\Lambda\}$. Then, $\widehat{\bm{\lambda}}$ is obtained by minimising the KL divergence from $q_{\lambda}(\bm\theta,\mathbf{x})$ to $p(\bm\theta,\mathbf{x}|\mathbf{y})$, which is equivalent to maximising the evidence lower bound (ELBO) function $\mathcal{L}(\bm{\lambda}) = E_{q_\lambda}\left[\log p(\mathbf{y},\mathbf{x}|\bm\theta)p(\bm\theta)-\log q_{\lambda}(\bm\theta,\mathbf{x})\right]$:
\begin{align}\label{eq:lambdahat}
\hat{\bm{\lambda}}=\operatornamewithlimits{argmin\,}_{\bm{\lambda}\in \Lambda}\text{KL}\left[ q_{\lambda}(\bm\theta,\mathbf{x})||p(\bm\theta,\mathbf{x}|\mathbf{y})\right]=\operatornamewithlimits{argmax\,}_{\bm{\lambda}\in \Lambda}\mathcal{L}(\bm{\lambda}).
\end{align}
VB methods that target the augmented posterior are much faster to implement relative to methods that approximate $p(\bm\theta|\mathbf{y})$ directly.

\subsection{Variational approximations for state space models}
This paper proposes a variational approximation of the form
\begin{align}\label{eq:vas}
    q_{\lambda}(\bm\theta,\mathbf{x})=q_{\lambda}(\bm\theta)q(\mathbf{x}|\mathbf{y},\bm\theta).
\end{align}
For the choice of $q_\lambda(\bm{\theta})$, we follow \cite{ong2018gaussian} and employ a $d$-dimensional Gaussian density with mean $\bm{\mu}$ and a covariance matrix with a factor structure representation $\Omega = BB^\top+\text{diag}(\bm{d}^2)$, where $B$ is a $d\times p$ matrix and $\bm{d}$ is a $d$-dimensional vector. The variational parameter vector is $\bm{\lambda} = (\bm{\mu}^\top,\bm{d}^\top,\text{vech}(B)^\top)^\top$, where the vech denotes the half vectorization of a rectangular matrix.

\citet{loaiza2022fast} show that the optimal choice of approximation for the latent states is $q(\mathbf{x}|\mathbf{y},\bm\theta) = p(\mathbf{x}|\mathbf{y},\bm{\theta})$, which guarantees exact integration of $\mathbf{x}$. However, the implementation of this approximation requires one to generate from $p(\mathbf{x}|\mathbf{y},\bm{\theta})$. This is computationally challenging or even infeasible for many state space model specifications, including nonlinear and high-dimensional state space models.

A faster approach which can be applied to a wider range of state space models, is to take $q(\mathbf{x}|\mathbf{y},\bm\theta) = q(\mathbf{x}|\bm{\theta})$. For instance, \cite{tan2018gaussian} and \citet{quiroz2018gaussian} take a multivariate Gaussian for $q(\mathbf{x}|\bm\theta)$ that does not condition on the data $\mathbf{y}$. 
\citet{frazier2022variational} show that this type variational approximations to latent states may lead to inferential and predictive inaccuracies.

This paper develops an accurate and fast variational Bayes method for the state space model in \eqref{eq:measurement}--\eqref{eq:state}, by proposing an approximation that can be expressed as $q(\mathbf{x}|\mathbf{y},\bm\theta)=q(\mathbf{x}|\mathbf{y})$. The proposed approximation is accurate due to the conditioning on $\mathbf{y}$. In addition, it is fast to implement as it does not directly condition on the parameter vector $\bm{\theta}$. The method is developed for models that have a closed-form measurement density $p(\mathbf{y}_t|\mathbf{x}_t,\bm\theta)$, and state transition density $p(\mathbf{x}_t|\mathbf{x}_{t-1},\bm\theta)$ that belongs to the exponential family of distributions. Generation from $p(\mathbf{x}|\mathbf{y},\bm{\theta})$ is not required. This makes our approach applicable to a wide range of different state space models. As will be discussed next, our approach is inspired by the literature on efficient importance sampling; hence we refer to it as Efficient VB.

\subsection{An efficient variational approximation to the states}\label{sec:approx}
We propose variational approximations to the states of the form
\begin{align}\label{eq:qxy}
    q(\mathbf{x}|\mathbf{y})=\prod_{t=1}^T q(\mathbf{x}_t|\mathbf{x}_{t-1},\mathbf{y},\bm\varphi),
\end{align}
where $\bm{\varphi}$ is an auxiliary parameter vector, which works as a proxy for $\bm{\theta}$. The conditional densities $q(\mathbf{x}_t|\mathbf{x}_{t-1},\mathbf{y},\bm\varphi)$ are written in terms of a transition kernel $k(\mathbf{x}_t,\mathbf{x}_{t-1}|\bm{a}_t,\bm\varphi)$ and an integration constant $\chi(\mathbf{x}_{t-1}|\bm{a}_t,\bm\varphi)=\int k(\mathbf{x}_t,\mathbf{x}_{t-1}|\bm{a}_t,\bm{\varphi}) d \mathbf{x}_t$:
\begin{align}\label{eq:vakernel}
    q(\mathbf{x}_t|\mathbf{x}_{t-1},\mathbf{y},\bm\varphi)=\frac{k(\mathbf{x}_t,\mathbf{x}_{t-1}|\bm{a}_t,\bm\varphi)}{\chi(\mathbf{x}_{t-1}|\bm{a}_t,\bm\varphi)},
\end{align}
where $\bm{a}_t$ is a vector of parameters  dependent on $\mathbf{y}$.

Denote $D[P||F]$ to be a divergence function between two distributions $P$ and $F$. The parameters $\bm{a} = (\bm{a}_1^\top,\dots,\bm{a}_T^\top)^\top\in A$ are calibrated so that $q(\mathbf{x}|\mathbf{y})$ accurately approximates $p(\mathbf{x}|\mathbf{y},\bm\varphi)$ as measured by $D$, that is
\begin{align}\label{eq:calibrate_a}
 \bm{a} = \argmin_{\tilde{\bm a}\in A} D\left[ \prod_{t=1}^T \frac{k(\mathbf{x}_t,\mathbf{x}_{t-1}|\tilde{\bm a}_t,\bm\varphi)}{\chi(\mathbf{x}_{t-1}|\tilde{\bm a}_t,\bm\varphi)}||p(\mathbf{x}|\mathbf{y},\bm{\varphi}) \right].
\end{align}

The optimization problem in \eqref{eq:calibrate_a} is similar to the one considered in efficient importance sampling \citep{richard2007efficient,koopman2015numerically}. Instead of sampling from the distribution of interest $p(\mathbf{x}|\mathbf{y},\bm\varphi)$, importance sampling replaces that distribution with an auxiliary distribution $q(\mathbf{x}|\mathbf{y})$. The parameters $\bm a$ are calibrated to minimize the variance of the ratio $p(\mathbf{x}|\mathbf{y},\bm\varphi)/q(\mathbf{x}|\mathbf{y})$.
We follow this approach, and solve \eqref{eq:calibrate_a} according to the steps proposed by \citet{richard2007efficient}.
Algorithm \ref{alg:vi2} in \citet{scharth2016particle} summarises this method for calibrating $\bm a$. 

The calibration of $\bm a$ in \eqref{eq:calibrate_a} is the most computationally expensive step in the VB optimization routine. However, if $q(\mathbf{x}|\mathbf{y})$ does not depend on $\bm\theta$, $\bm a$ does not have to be calibrated at each iteration of VB. Hence, we set $\bm\varphi$ to be a parameter vector close to $\bm\theta$ rather than $\bm\theta$ itself, and only update $\bm\varphi$ and $\bm a$ every 200 steps. The numerical experiments in Section~\ref{sec:simulation} and the empirical applications in Section~\ref{sec:skellam} and \ref{sec:application} demonstrate that this does not hinder the accuracy of the approach. 

Since the transition density belongs to the exponential family of distributions, it can be written as 
\begin{equation}
p(\mathbf{x}_t|\mathbf{x}_{t-1},\bm\varphi) = h(\mathbf{x}_t)g(\mathbf{x}_{t-1},\bm{\varphi})\exp\left(\bm{\eta}(\mathbf{x}_{t-1},\bm{\varphi})^\top\bm{T}(\mathbf{x}_t)\right),
\end{equation}
where $\bm{T}(\mathbf{x}_t)$ denotes a vector of sufficient statistics. 
We select the transition kernel $k(\mathbf{x}_t,\mathbf{x}_{t-1}|\bm a_t,\bm\varphi)$ to be
\begin{align}\label{eq:kernel}
    k(\mathbf{x}_t,\mathbf{x}_{t-1}|\bm a_t,\bm\varphi) = \exp\left(\bm{a}_t^\top\bm{T}(\mathbf{x}_t)\right)p(\mathbf{x}_t|\mathbf{x}_{t-1},\bm\varphi).
\end{align}
The properties of the exponential family in combination with this choice of kernel guarantees the practical applicability of the proposed method because calibration of $q(\mathbf{x}|\mathbf{y})$ can be implemented via an algorithm that involves a fast recursive sequence of linear regressions. 
This algorithm is feasible because $ q(\mathbf{x}|\mathbf{y})$ is easy to generate from and the integration constant $\chi(\mathbf{x}_{t-1}|\bm{a}_t,\bm\varphi)$ can be evaluated, as shown in Appendix~\ref{app:calibration}. 

The suggested approximation $q(\mathbf{x}|\mathbf{y})$ differs from existing Gaussian approximations in three ways. First, it allows for choices of exponential kernels that do not lead to a Gaussian approximation on the states. Second, note that while the approximation does not directly condition $\bm{\theta}$, it can condition on the proxy parameter $\bm{\varphi}$ in a non-linear fashion. That is, even if a Gaussian kernel was considered, the Efficient VB method has the ability to capture non-linear relationships between $\mathbf{x}$ and $\bm{\theta}$ via the proxy parameter, which the Gaussian approximation cannot. Third, the recursive nature of $q(\mathbf{x}|\mathbf{y})$ accurately captures the dependence structure of the exact conditional posterior $p(\mathbf{x}|\mathbf{y},\bm{\theta})$. 


\subsection{Stochastic gradient ascent}
We solve the optimization problem in \eqref{eq:lambdahat} using SGA methods. SGA calibrates the variational parameter by iterating over 
\begin{equation}\label{eq: lambda_iter}
    \bm{\lambda}^{[j+1]} = \bm{\lambda}^{[j]} + \bm{\rho}^{[j]}\circ\widehat{\nabla_\lambda \mathcal{L}\left(\bm{\lambda}^{[j]}\right)}, 
\end{equation}
until convergence is achieved. The vector $\bm{\rho}^{[j]}$ contains the so called ``learning parameters'', which we set according to the ADADELTA approach in \cite{zeiler2012adadelta}. The vector $\widehat{\nabla_\lambda \mathcal{L}\left(\bm{\lambda}^{[j]}\right)}$ is an unbiased estimate of the gradient of the ELBO 
evaluated at $\bm{\lambda}^{[j]}$. 

Any draw from $q_\lambda(\bm{\theta})$ can be expressed as $\bm{\theta} = \bm{\theta}(\bm{\varepsilon},\bm{\lambda})$, where $\bm{\varepsilon}\sim f_{\varepsilon}$ and $f_{\varepsilon}$ is a distribution that does not depend on $\bm{\lambda}$. Using the re-parametrization trick in \citet{kingma2013auto} the ELBO gradient is written as 
\begin{align}\label{eq:gradient_elbo}
    \nabla_\lambda\mathcal{L}(\bm{\lambda}) = E_{q(\mathbf{x}|\mathbf{y}),f_\varepsilon}\left[\frac{\partial \bm{\theta}}{\partial\bm{\lambda}}^\top\left[\nabla_\theta\log p(\mathbf{y},\mathbf{x}|\bm\theta)p(\bm\theta)-\nabla_\theta\log q_\lambda(\bm\theta)\right] \right],
\end{align}
where the expectation is taken with respect to $q(\mathbf{x}|\mathbf{y})$ and $f_\varepsilon$. The gradient $\nabla_\theta\log p(\mathbf{y},\mathbf{x}|\bm\theta)p(\bm{\theta})$ is model specific. The expressions $\partial\bm{\theta} /\partial\bm{\lambda}$ and $\nabla_\theta \log q_\lambda(\bm{\theta} )$ are provided in \cite{ong2018gaussian}.
At each SGA iteration $[j]$, we calculate a sample estimate of \eqref{eq:gradient_elbo} based on only one draw for both $\mathbf{x}$ from $q(\mathbf{x}|\mathbf{y})$ and $\bm\varepsilon$ from $f_\varepsilon$. Note that $q(\mathbf{x}|\mathbf{y})$ only depends on the parameters $\bm\varphi$ and $\bm a$, which are updated by setting $\bm\varphi = \bm\mu^{[j]}$ and re-calibrating $\bm a$ as in \eqref{eq:calibrate_a}.
The VB estimation routine is summarized in Algorithm~\ref{alg:vi}. 

\begin{algorithm}
    \begin{algorithmic}[1]
        \State{Initialize $\bm\lambda^{[0]}$ and set iteration $j=0$.}
        \While{no convergence of ELBO}
        \State{Set $j=j+1$.}
        \If{$j+199$ is a multiple of 200}
        \State{Set $\bm\varphi=\bm\mu^{[j]}$.}
        \State{Solve $\bm a = \argmin_{\tilde{\bm a}\in A} D\left[ \prod_{t=1}^T \frac{k(\mathbf{x}_t,\mathbf{x}_{t-1}|\tilde{\bm a}_t,\bm\varphi)}{\chi(\mathbf{x}_{t-1}|\tilde{\bm a}_t,\bm\varphi)}||p(\mathbf{x}|\mathbf{y},\bm{\varphi}) \right]$.}
        \State{Set $q(\mathbf{x}|\mathbf{y})=\prod_{t=1}^Tq(\mathbf{x}_t|\mathbf{x}_{t-1},\mathbf{y},\bm\varphi)$.}
        \EndIf
        \State{Draw $\bm\varepsilon^{[j]}$ from $f_\varepsilon$ and set $\bm\theta^{[j]} = h(\bm\varepsilon^{[j]},\bm\lambda^{[j]})$.}
        \State{Draw $\mathbf{x}^{[j]}$ from $q(\mathbf{x}|\mathbf{y})$.}
        \State{Compute $\widehat{\nabla_\lambda \mathcal{L}\left(\bm{\lambda}^{[j]}\right)}=\left.\frac{\partial \bm{\theta}}{\partial\bm{\lambda}}^\top\left[\nabla_\theta\log p(\mathbf{y},\mathbf{x}|\bm{\theta})p(\bm{\theta})-\nabla_\theta\log q_\lambda(\bm{\theta})\right]\right|_{\bm\theta=\bm{\theta}^{[j]},\bm\lambda=\bm{\lambda}^{[j]},\mathbf{x}=\mathbf{x}^{[j]}}$.}
        \State{Update $\bm{\rho}^{[j]}$ via ADADELTA.}
        \State{Set $\bm{\lambda}^{[j+1]} = \bm{\lambda}^{[j]} + \bm{\rho}^{[j]}\circ\widehat{\nabla_\lambda \mathcal{L}\left(\bm{\lambda}^{[j]}\right)}$.}
        \EndWhile
    \end{algorithmic}
    \caption{Efficient VB algorithm}
    \label{alg:vi}
\end{algorithm}

Since line 6 in Algorithm \ref{alg:vi} takes the most computation time, $\bm\varphi$ and $\bm a$ are only updated every 200 steps, as discussed in Section~\ref{sec:approx}. Appendix~\ref{app:calibration} provides a detailed algorithm together with additional details on how line 6 is implemented. 
\section{Numerical experiments}\label{sec:simulation}
This section presents numerical experiments to assess the accuracy and the computational costs of the proposed VB approach in a stochastic volatility model. This is a state space model that is widely used in economics, and that can also be estimated by MCMC and existing VB methods. This allows for the investigation of the properties of our method relative to benchmark methods for varying sample sizes. 

\subsection{Stochastic volatility model}
The stochastic volatility model is defined as 
\begin{eqnarray}
p(y_t|x_t)& = &\phi_1(y_t;0,e^{x_t}),\nonumber\\ 
p(x_t|x_{t-1},\bm{\theta})& = &\phi_1(x_t;\bar{x}+\rho(x_{t-1}-\bar{x}),\sigma^2),\,
\label{eq:sv}
\end{eqnarray}
where $\phi_1(x;\mu,s^2)$ denotes the univariate Gaussian density function with mean $\mu$ and variance $s^2$,  $\bm{\theta} = (\bar{x},\rho,\sigma)'$ are the parameters of the model, and $x_t$ denotes the latent log-variances of the time series process at time $t$. We generate $T = 4000$ observations from the model in \eqref{eq:sv} with the true parameter values set as $\bar{x}_0 = -1.3$, $\rho_0 = 0.95$, and $\sigma_0 = 0.3$. 

The objective of the experiments is to assess the accuracy and computational costs of different VB methods in approximating the augmented posterior
\begin{equation}
p\left(\bm{\theta},\xvec|\bm{y}\right)\propto p\left(\bm{\theta}\right)p(x_1|\bm{\theta})\prod_{t=2}^{T} p\left(y_t|x_t\right)p\left(x_t|x_{t-1},\bm{\theta}\right),
\label{eq:ssmpost}
\end{equation}
for varying sample sizes, where $p(x_1|\bm{\theta})=\phi_1(x_1;\frac{\bar{x}}{1-\rho},\frac{\sigma^2}{1-\rho^2})$. Here, $\xvec=(x_1,\ldots,x_T)^\top$ and $p\left(\bm{\theta}\right) = p(\bar{x})p(\rho)p(\sigma)$ is the prior density for $\theta$, with $p(\bar{x})=N(0,1000)$, $p(\rho)=\text{Uniform}(0,0.995)$, and $p(\sigma)=\text{Inverse-Gamma}(1.001,1.001)$. VB is implemented by transforming all the parameters to the real line so that $\rho = 0.995/(1+\exp(-\kappa))$, and $\sigma = \exp(c/2)$.

The MCMC sampler generates from the exact posterior \eqref{eq:ssmpost}. Appendix~\ref{A:sv} outlines the steps of the MCMC algorithm. Because of the low computational costs of implementing MCMC, we can estimate the model for multiple sample sizes. Specifically, we estimate the model using $T = 1,10,20,\dots,4000$ observations. These results provide insights into the accuracy and computational costs of the different VB methods in small and large samples. 

\subsection{Variational approximations}
The stochastic volatility model allows for the construction of the efficient variational approximation in  \eqref{eq:qxy}. 
Since the state transition density is Gaussian, set $\bm T( x)=(x,x^2)^\top$ and  $\bm a_t = (b_t,c_t)^\top$ with $b_t$ and $c_t$ both scalars.
Denote $\bm\varphi = (\bar{x}_\varphi,\rho_\varphi,\sigma_\varphi)$. The approximation to the states $q(\mathbf{x}|\mathbf{y}) = \prod_{t=1}^Tq(x_t|x_{t-1},\mathbf{y},\bm\varphi)$ is a product of Gaussian densities such that $q(x_t|x_{t-1},\mathbf{y},\bm\varphi) = \phi_1(x_t;\mu_t,\sigma_t^2)$ with $\sigma_{t} = (\sigma_\varphi^{-2}-2c_{t})^{-1/2}$, $\mu_{t} = \sigma_{t}^2\left[b_{t}+\frac{\bar{x}_\varphi+\rho_\varphi(x_{t-1}-\bar{x}_\varphi)}{\sigma_\varphi^2}\right]$, and normalising constant
\begin{align}
  \chi(x_{t-1}|\bm{a}_t,\bm\varphi) =  \exp\left[ \frac{1}{2}\log \frac{\sigma_{t}}{\sigma_\varphi}+\frac{1}{2}\frac{\mu_{t}^2}{\sigma_{t}^2}-\frac{1}{2}\frac{(\bar{x}_\varphi+\rho_\varphi(x_{t-1}-\bar{x}_\varphi))^2}{\sigma_\varphi^2}\right].
\end{align}
Because generation of a random draw from $ q(\mathbf{x}|\mathbf{y}) $ is fast, so is line 10 in Algorithm 1.

In addition, we also implement the Hybrid VB  approach proposed by \citet{loaiza2022fast} and a Gaussian variational approximation. Hybrid VB requires draws from the conditional density $p(\mathbf{x}|\mathbf{y},\bm\theta)$, which can be generated using the filtering steps discussed in \citet{kim1998stochastic}. Gaussian VB takes $q(\mathbf{x}|\bm\theta)=\phi_T(\mathbf{x},\bm\mu_x,C_xC_x^\top)$ to be a $T$-dimensional multivariate Gaussian density, where the Cholesky factor $C_x$ is a lower triangular matrix with three non-negative bands. In all three methods we use a Gaussian approximation with a factor covariance matrix for $\bm{\theta}$ and set the number of factors to one. The gradient expressions required for the implementation of all three VB methods are provided in Appendix~\ref{A:sv}. We run the VB algorithms for a total of 10,000 iterations. MCMC is implemented using a burn-in sample of size 10,000 an inference sample of size 10,000.

\subsection{Results}

\subsubsection{Accuracy of the posterior distribution for the states }
\begin{figure}[tb!]
\caption{Posterior mean of the states in the numerical experiment with $T=500$}
\centering
\includegraphics*[width=\textwidth]{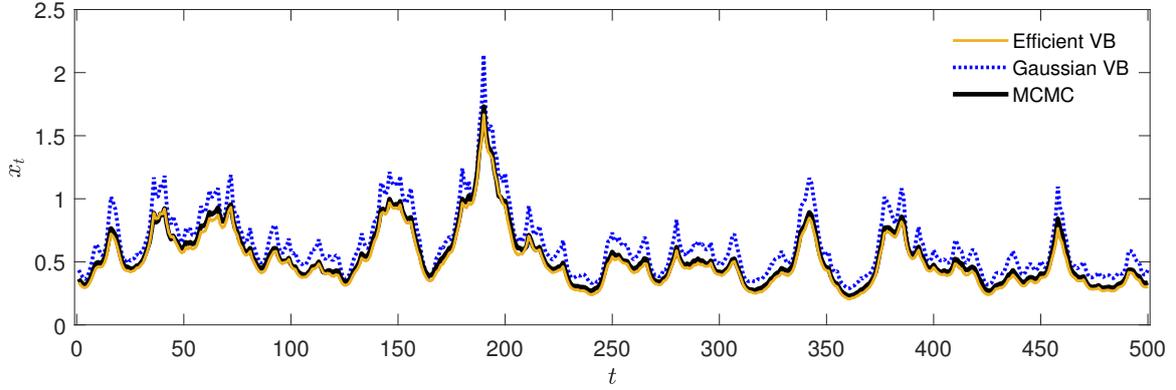}\\[3mm]
\begin{flushleft}
This figure shows the posterior mean in the numerical experiment with an estimation sample of 500 observations, for Efficient VB, Gaussian VB, and MCMC. These are indicated by the solid yellow, dotted blue, and solid black line, respectively. 
\end{flushleft}
\label{fig:SVstates}
\end{figure}

First, we assess the accuracy of the posterior distribution for the states in $\mathbf{x}$ in a small sample with $T =500$ observations. Figure~\ref{fig:SVstates} shows the Efficient VB, Gaussian VB, and MCMC posterior means of the states. Since Hybrid VB uses the exact conditional density of the states in its variational approximation, its posterior mean for the states is very similar to that of MCMC and is not included in the figure. Although Efficient VB uses an approximation to the conditional state density, its posterior mean is almost identical to the posterior mean of MCMC over the whole sample period. This is not the case for the posterior mean of Gaussian VB. The Gaussian VB posterior means overestimate the states compared to the posterior means from MCMC in almost each time period. 

\begin{figure}[tb!]
\caption{Posterior correlations between the states in the numerical experiment with $T=500$}
\centering
\includegraphics*[width=\textwidth]{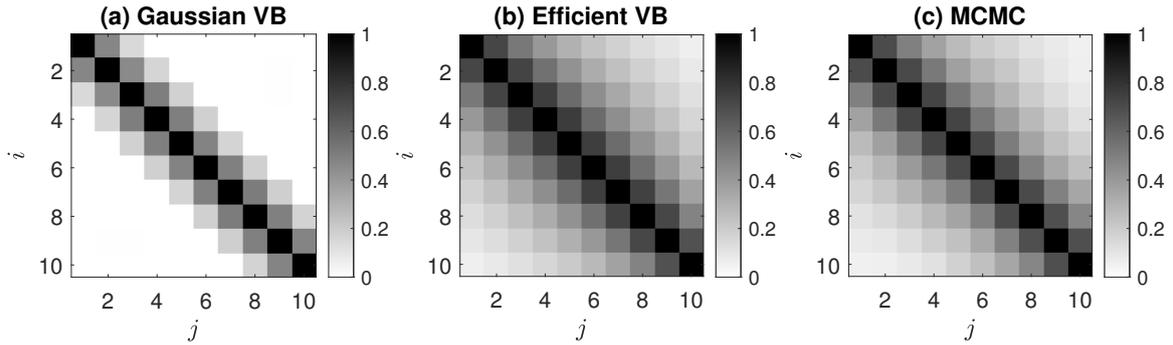}\\[3mm]
\begin{flushleft}
This figure shows the posterior correlations between $x_{100+i}$ and $x_{100+j}$ with $i,j=1,\dots,10$, in the numerical experiment with an estimation sample of 500 observations, for Gaussian VB, Efficient VB, and MCMC.
\end{flushleft}
\label{fig:SVcor}
\end{figure}

Additionally, we analyse the posterior dependence structure of the states. Panels (a) to (c) in Figure~\ref{fig:SVcor} show the  posterior correlations between $x_{100+i}$ and $x_{100+j}$ for $i,j=1,\dots,10$, for Gaussian VB, Efficient VB and MCMC, respectively. Gaussian VB underestimates most pairwise posterior correlations. Panel (a) in Figure~\ref{fig:SVcor} shows that only the first and second order posterior correlations are nonzero, while the MCMC posterior correlations are positive up to at least the tenth order. The posterior correlations of Efficient VB and MCMC do not show any differences. We find the same patterns across different time periods and across longer time samples. 

The differences in accuracy of the posterior distribution for the states can be explained by the choice of variational approximation for the states. Gaussian VB calibrates $q(\mathbf{x}|\bm\theta)$, which is far from the ideal distribution  $p(\mathbf{x}|\mathbf{y},\bm\theta)$, as it does not condition on $\mathbf{y}$. On the other hand, Efficient VB uses $q(\mathbf{x}|\mathbf{y})$, which directly conditions on $\mathbf{y}$. Figures~\ref{fig:SVstates} and \ref{fig:SVcor} suggest that when it comes to accurately representing the posterior distribution of the states, it is important to directly condition on the data. While estimation of the posterior of the states is not always the target of Bayesian analysis, accurate estimation of this posterior is critical to obtaining accurate variational approximations to the target posterior $p(\bm{\theta}|\mathbf{y})$. We demonstrate this in the next section.

\subsubsection{Accuracy of the posterior distribution for the parameters}
\begin{figure}[tb!]
\caption{Posterior parameter distributions in the numerical experiment with $T=500$}
\centering
\includegraphics*[scale=0.7]{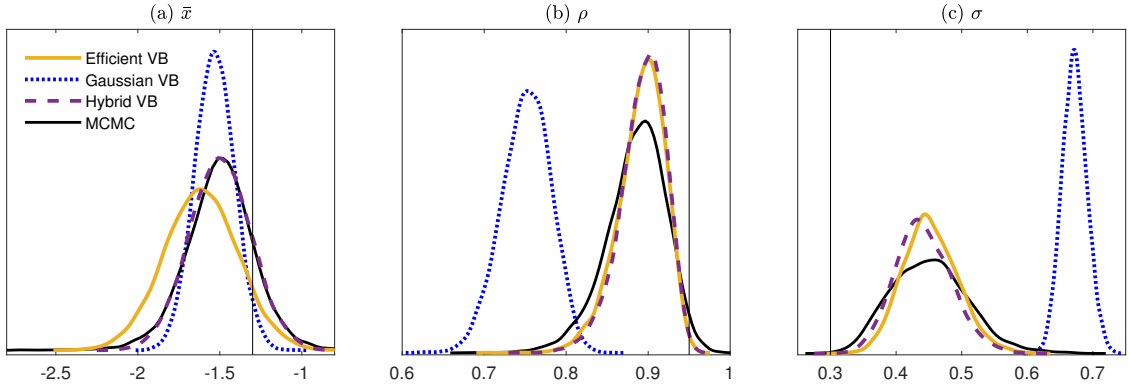}\\[3mm]
\begin{flushleft}
This figure shows the posterior parameter distributions in the numerical experiment with an estimation sample of 500 observations, for Efficient VB, Gaussian VB, Hybrid VB, and MCMC. These are indicated by the solid yellow, dotted blue, dashed purple, and solid black line, respectively. Panel (a) shows the posterior distribution for $\bar{x}$, Panel (b) for $\rho$, and Panel (c) for $\sigma$. The vertical lines indicate the true values in the data generating process.
\end{flushleft}
\label{fig:SVsmall}
\end{figure}

Second, we assess the accuracy of the posterior distribution for the parameters. Figure~\ref{fig:SVsmall} shows the posterior parameter distributions for Efficient VB (solid yellow), Gaussian VB (dotted blue), Hybrid VB (dashed purple), and MCMC (solid black) in a small sample with 500 observations. The posterior of Efficient VB is close to the exact posterior of MCMC for all three parameters. The small differences between these posteriors can be summarized as a slight change in location for $\bar{x}$ and underestimation of the variance of the posteriors of $\rho$ and $\sigma$, which is a well-known property of VB. Hence, Hybrid VB also underestimates the posterior variance, but is slightly more accurate in the posterior locations. However, Gaussian VB only produces an accurate approximation for $\bar{x}$. 

\begin{figure}[tb!]
\caption{Posterior parameter distributions for different sample sizes}
\centering
\includegraphics*[width=\textwidth]{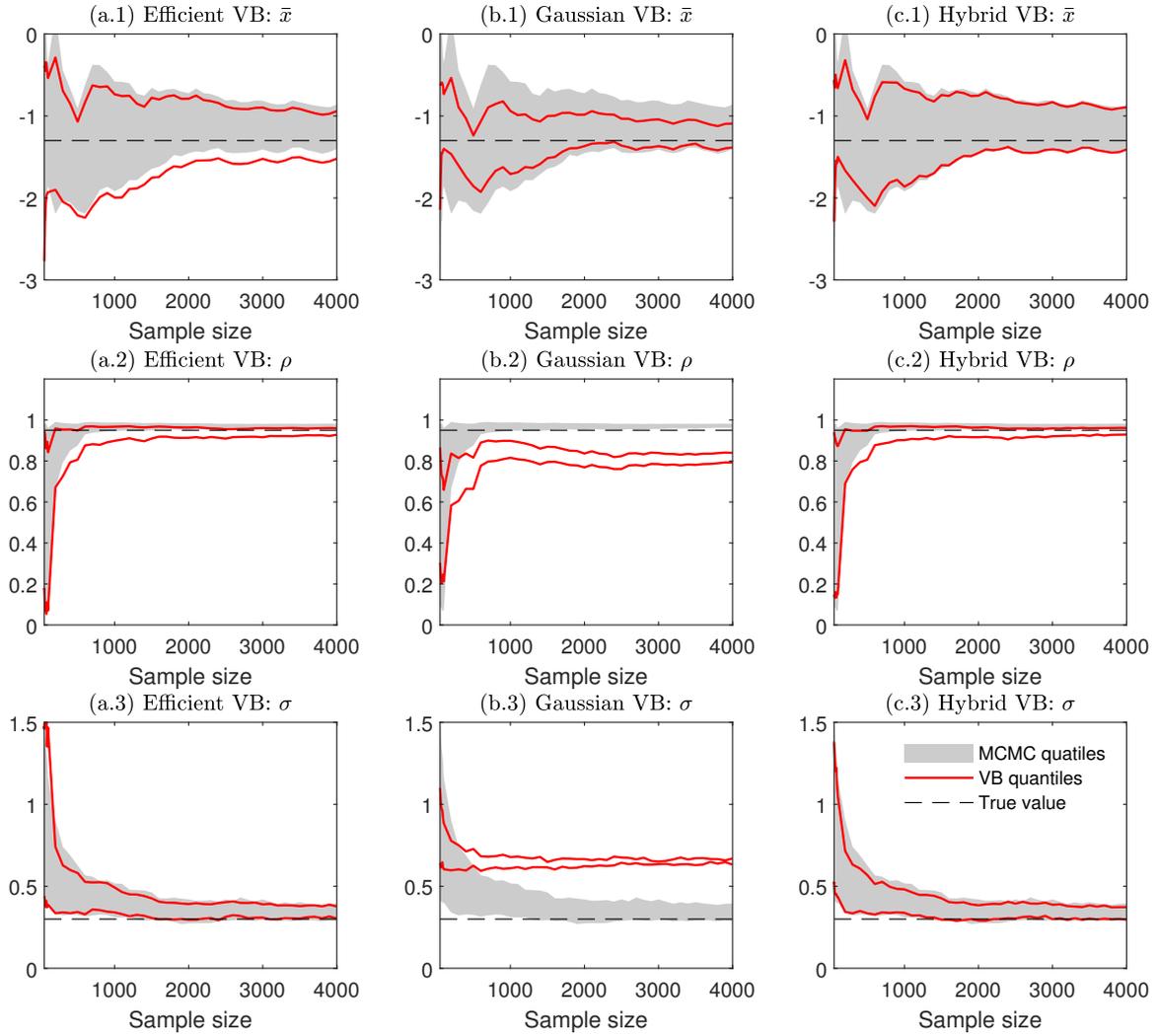}\\[3mm]
\begin{flushleft}
This figure shows the 0.5\% and 99.5\% quantiles of the posterior parameter distributions in the numerical experiments with different estimation samples. The columns correspond to Efficient VB, Gaussian VB, and Hybrid VB, whose quantiles are indicated by solid red lines and compared to the quantiles of MCMC indicated by gray areas. The rows correspond to the parameters $\bar{x}$, $\rho$, and $\sigma$. The horizontal dashed lines indicate the true values in the data generating process.
\end{flushleft}
\label{fig:SVlarge}
\end{figure}

Additionally, we assess how the accuracy of the approximations changes with the sample size. The red lines in Panels (a.1) to (a.3) in  Figure~\ref{fig:SVlarge} show the 99\% posterior intervals for all the three parameters using Efficient VB. The shaded areas correspond to the MCMC posterior intervals. The x-axis indicates the sample size used for estimation. We find that the posterior intervals of Efficient VB are similar to those of MCMC, and the accuracy does not seem to be affected by the sample size. The posterior interval of Efficient VB concentrates to a location close to the true values with increasing sample size. Panels (c.1) to (c.3) in the figure show that the posterior intervals of Hybrid VB are more accurate for $\bar{x}$, but are not that different from Efficient VB for $\rho$ and $\sigma$. Panels (b.1) to (b.3) show that Gaussian VB is less accurate for all parameters and all sample sizes under consideration. Moreover, as the sample size increases its posteriors do not concentrate close to the true parameter values. This behaviour is likely to be related to the approximation errors to the posterior of the states that are exhibited by Gaussian VB.

\subsubsection{Computation time}
\begin{figure}[tb!]
\caption{Estimation time in the numerical experiments}
\centering
\includegraphics*[scale=0.7]{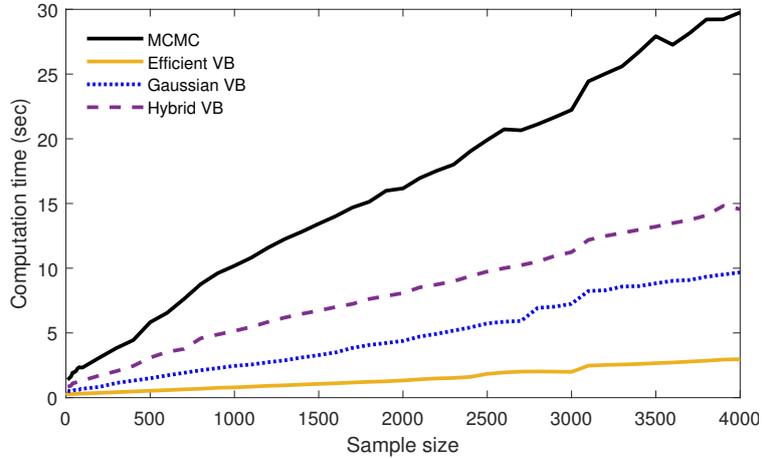}\\[3mm]
\begin{flushleft}
This figure shows the estimation time in seconds in the numerical experiments with different sample sizes, for Efficient VB, Gaussian VB, Hybrid VB, and MCMC. These are indicated by the solid yellow, dotted blue, dashed purple, and solid black line, respectively. 
\end{flushleft}
\label{fig:ComputationTimes}
\end{figure}

Third, we compare the computational costs of the different estimation methods. Figure~\ref{fig:ComputationTimes} shows the estimation time with increasing sample sizes for Efficient VB, Gaussian VB, Hybrid VB, and MCMC. Efficient VB is substantially faster than the other methods: more than six times faster than MCMC, but also three times faster than Hybrid VB and more than twice as fast as Gaussian VB with a sample size of 4000 observations. Hence, Efficient VB is both more accurate and faster relative to Gaussian VB. Although the differences between the methods are smaller for smaller sample sizes, the ordering in the computational costs remains the same. 

The substantial reduction in computational costs by Efficient VB can be explained by two properties of its variational approximation to the states. First, Efficient VB does not require computationally costly filtering steps to calibrate its approximation to the states. This explains the computational gains relative to Hybrid VB, which samples from the exact conditional density of the states by forward filtering and backward smoothing. Second, the number of variational parameters in Efficient VB depends only on the dimension of $\bm\theta$. On the other hand, the number of variational parameters in Gaussian VB also depends on the dimension of $\mathbf{x}$, making the required gradient and matrix operations computationally costly, especially as the sample size $T$ increases.

The computation time of Efficient VB is approximately equally distributed between three steps of Algorithm~\ref{alg:vi}: the calibration of the approximation to the states in line 6, the generation of the states in line 10 and the estimation of the gradient in line 11. The generation of the states is  efficient since it does not require filtering steps, and so is evaluation of the gradient as its dimension is not affected by $T$. On the other hand, calibration of $q(\mathbf{x}|\mathbf{y})$ can be costly, since it involves a recursive sequence of linear regressions that increases in the number of observations. As a result, we only run step 6 of the algorithm every 200 iterations. The results in Figure \ref{fig:SVlarge} indicate that this choice of update frequency does not hinder the accuracy of the model. This is also corroborated by the sensitivity analysis presented in Appendix~\ref{A:numerical}, which demonstrates that more frequent updates of $q(\mathbf{x}|\mathbf{y})$ do not produce higher accuracy of the variational approximation for either small or large sample sizes. We find similar results for the state space models with high-dimensional state vectors discussed in Section~\ref{sec:skellam} and \ref{sec:application}. 
\section{The multivariate Skellam stochastic volatility model}\label{sec:skellam}
To illustrate Efficient VB with real data, we first fit a multivariate Skellam stochastic volatility model for high-frequency tick-by-tick discrete price changes. The Skellam stochastic volatility model is known to be a challenging model to estimate. The model is introduced by \citet{koopman2017intraday} in a univariate setting. Since their inferential approach is not feasible for a multivariate setting, multivariate Skellam models have been proposed outside the state space framework \citep{koopman2018dynamic,catania2022dynamic}. We extend the model to multiple assets by modelling each asset with independent Skellam distributions conditional on a multivariate transition equation that allows for volatility clustering and dependence between assets. 

This application of Efficient VB to a model with a non-linear measurement equation and a multivariate state vector demonstrates that our approach is also fast and accurate in complex state space models. In contrast, 
the VB method of \cite{tran2017variational}, which uses particle filtering, is computationally impractical with a multivariate state vector: a large number of particles would be required to produce a low variance estimate of the gradient required for its implementation. Hybrid VB is also infeasible in this model as it requires generation from the full-conditional $p(\mathbf{x}|\mathbf{y},\bm{\theta})$, for which no computationally efficient method is currently available.

The data is obtained from the New York Stock Exchange (NYSE) trades and quotes database, from which we consider trades at the NYSE of the four U.S. companies studied in \cite{koopman2017intraday}: Walmart Stores Inc. (WMT), The Coca-Cola Company (KO), JPMorgan Chase \& Co. (JPM), and Caterpillar Inc. (CAT). We consider price changes at a 15 second frequency in one of the most volatile weeks in the last decade: June 8th to June 12th in 2020. After applying the data cleaning steps P1, T1, T2 and T3 of \cite{barndorff2009realized}, prices are constructed with the previous-tick method. Since trading hours run from 9:30:00am to 4:00:00pm, there are 1560 price observations at the 15 second frequency, which after differencing result in 1559 price changes. The final sample includes 5$\times$1559 = 7795 price changes for the week under consideration.

\subsection{The model}
Let $y_{i,t}\in \mathbb{Z}$ denote the price change of asset $i=1,\dots,N$ at time $t$. We assume that these price changes follow a zero-inflated Skellam distribution as proposed by \citet{catania2022dynamic}: 
\begin{equation}\label{eq:zi_skellam}
p(y_{i,t}|x_{i,t},\bm{\theta}) = \kappa_i \delta_{0}(y_{i,t})+(1-\kappa_i)\exp\left(-\sigma^2_{i,t}\right)\mathcal{I}_{|y_{i,t}|}(\sigma^2_{i,t}),
\end{equation}
where $\delta_{0}(y_{i,t})$ is a Dirac point mass function at $0$ and $\exp\left(-\sigma^2_{i,t}\right)\mathcal{I}_{|y_{i,t}|}(\sigma^2_{i,t})$ denotes the Skellam probability mass function, with $\mathcal{I}_{|y_{i,t}|}(\cdot)$ the modified Bessel function of order $|y_{i,t}|$. To account for different amounts of zero returns, $\kappa_i$ governs the zero-inflation. The stochastic volatility is defined as $\sigma^2_{i,t} = \exp(s_{i,t}+x_{i,t})$, where $s_{i,t}$ captures the seasonal variation in intraday volatility, and $x_{i,t}$ is a state variable that captures the non-seasonal variation. 

We assume that the price changes are conditionally independent and define the  measurement equation
\begin{align}
p(\yvec_t|\xvec_t,\bm{\theta}) = \prod_{i=1}^N p(y_{i,t}|x_{i,t},\bm{\theta}),\label{Eq:measure}
\end{align}
with $\yvec_t=(y_{1,t},y_{2,t},\ldots,y_{N,t})^\top$. The state vector of the model is $\xvec_t=(x_{1,t},x_{2,t},\ldots,x_{N,t})^\top$ for which we assume the transition equation
\begin{align}
    p(\xvec_t|\xvec_{t-1},\bm{\theta}) = \phi_{N}\left(\xvec_t;\bar{\bm x}+\Omega\xvec_{t-1},\Sigma\right),\label{Eq:state}
\end{align}
where $\bar{\bm x}=(\bar{x}_1,\dots,\bar{x}_N)^\top$, $\Omega$ is a diagonal matrix with elements $\bm\omega=(\bm\omega_{1},\dots,\bm\omega_N)^\top$, and $\Sigma =(LL^\top)^{-1}$ a covariance matrix. Persistence in volatility is captured by $\bm\omega $ and cross-sectional dependence between the volatilities of the stocks by $\Sigma$. The full parameter vector of this model is  $\bm\theta=(\bm\kappa^\top,\bar{\bm x}^\top,\bm\omega^\top,\text{vech}(L)^\top)^\top$, where $\bm\kappa=(\kappa_1,\dots,\kappa_N)^\top$.

The seasonal component is constructed as a cubic spline function $s_{i,t}=\tilde{W}_t^\top \bm{\beta}_i$ with basis $\tilde{W}_t$. The basis functions are constructed in two steps. First, we use the steps in section 9.1.1 of \cite{greenberg2012introduction} with four knots located at times 9:30:30am, 10:00:30am, 12:30:00pm and 4:00:00pm to construct the basis $W_t = (W_{1,t},\dots,W_{4,t})^\top$. Second, we apply the transformation in \cite{harvey1993forecasting} to form the zero sum basis vector $\tilde{W}_t = (\tilde{W}_{1,t},\tilde{W}_{2,t},\tilde{W}_{3,t})^\top$, where $\tilde{W}_{j,t}={W}_{j,t}-{W}_{4,t}\frac{\bar{W}_j}{\bar{W}_4}$ and $\bar{W}_j=\sum_{t=1}^{1559}{W}_{j,t}$. This second step allows us to identify $\bar{x}_i$ by ensuring that $\sum_t \tilde{W}_t^\top \bm{\beta}_i=0$.

The augmented posterior of $\bm\theta$ and $\xvec$ conditional on $\yvec$ is
\begin{align}\label{eq:post_skellam}
p(\thetavec,\xvec|\yvec) \propto& p(\yvec|\xvec,\thetavec)p(\xvec|\thetavec)p(\thetavec) \\
 =&
\prod_{t=1}^{T}p(\yvec_t|\xvec_t,\bm{\theta}) \phi_{N}\left(\xvec_t;\bar{\bm x}+\Omega\xvec_{t-1},\Sigma\right)
p(\thetavec),
\end{align}
where $\xvec_0 = (\log\text{var}(y_1),\dots,\log \text{var}(y_N)))^\top$. The prior density for $\thetavec$ is specified as $p(\thetavec)= p(L)\prod_{i=1}^Np(\kappa_i)$ $p(\bar{x}_i)p(\omega_i)p(\beta_i)$ with $\kappa_i\sim \text{Uniform}(0,1)$,  $\bar{x}_i\sim N(0,100)$, $\omega_i\sim \text{Uniform}(0,1)$, $\bm{\beta}_i\sim N(\bm{0}_3,100I_3)$ and $p(L) = |LL^\top|^{-\frac{N+1}{2}}\prod_{i=1}^NL_{i,i}(N+1-i)$, which implies a Jeffrey's prior on $\Sigma$.

\subsection{Variational approximation}
The augmented posterior of the model admits an approximation as proposed in \eqref{eq:qxy}. 
%
Since the state transition density is a multivariate Gaussian, $\bm T( \mathbf{x}_t)=(\mathbf{x}_{t}^\top,\text{vec}(\mathbf{x}_{t}\mathbf{x}_{t}^\top))^\top$. We define the vector of kernel parameters to be $\bm a_{t} = (\mathbf{b}_{t},\text{vec}(C_{t}))^\top$ with $\bm{b}_{t}$ an $N$-dimensional vector and $C_{t} = \text{diag}(\bm{c}_{t})$ specified as a diagonal matrix for computational efficiency, where $\bm{c}_{t}$ is an $N$-dimensional vector. 

The approximation to the states $q(\mathbf{x}|\mathbf{y}) = \prod_{t=1}^Tq(\mathbf{x}_{t}|\mathbf{x}_{t-1},\mathbf{y},\bm\varphi)$ is a product of Gaussian densities such that $q(\mathbf{x}_{t}|\mathbf{x}_{t-1},\mathbf{y},\bm\varphi) = \phi_{N}(\mathbf{x}_{t};\bm\mu_{t},V_{t})$ with $\bm{\mu}_{t} = V_{t} \left(\bm{b}_{t}+\Sigma^{-1}\left(\bar{\mathbf{x}}+\Omega\mathbf{x}_{t-1}\right)\right)$ and $V_{t} = \left(\Sigma^{-1}-2C_{t}\right)^{-1}$. The integration constant of the transition kernel equals
\begin{align}
  \chi(\mathbf{x}_{t-1}|\bm a_{t},\bm\varphi)
  &=  \exp\left[ \frac{1}{2}\log \frac{|V_{t}|}{|\Sigma|}+\frac{1}{2}\bm{\mu}_{t}^\top V_{t}^{-1}\bm{\mu}_{t}-\frac{1}{2}\left(\bar{\mathbf{x}}+\Omega\mathbf{x}_{t-1}\right)^\top \Sigma^{-1}\left(\bar{\mathbf{x}}+\Omega\mathbf{x}_{t-1}\right) \right]. \notag
\end{align}
While not made explicit in the notation, the parameters $\bar{\mathbf{x}}$, $\Omega$ and $\Sigma$ in the approximation are determined by the proxy parameter vector $\bm\varphi$. 
Since we assume $C_{t}$ to be diagonal, only the kernel parameters $\bm{b}_{t}$ and $\bm{c}_{t}$ have to be calibrated. Hence, $\bm{\gamma}_{t}^\top\bm{T}(\mathbf{x}_{t}^{[s]})$ boils down to $\bm{\tilde{\gamma}}_{t}^\top(\mathbf{x}_{t}^\top,(\mathbf{x}_{t}^2)^\top )^\top$ in Algorithm~\ref{alg:vi2} in Appendix~\ref{app:calibration}.

In addition to our method, we also consider the Gaussian VB approach. Gaussian VB takes $q(\mathbf{x}|\bm\theta)=\phi_T(\mathbf{x},\bm\mu_x,C_xC_x^\top)$ to be a $T$-dimensional multivariate Gaussian density, where the Cholesky factor $C_x$ is a lower triangular matrix with three non-negative bands. Both VB methods use a Gaussian approximation with a factor covariance matrix for $\bm{\theta}$ and set the number of factors to two. To judge the accuracy of the approximations we also implement PMCMC. Appendix~\ref{A:skellam} presents further implementation details of all methods in this example. PMCMC is implemented using a burn-in sample of size 15,000, an inference sample of size 15,000, and a total of 1,000 particles. We run both VB algorithms for a total of 15,000 iterations. The ELBO figures in Appendix~\ref{A:skellam_results} shows that the number of VB iterations suffices to achieve convergence.

\subsection{Results}

\subsubsection{Univariate case}
We assess the accuracy of Efficient VB by comparing its posterior parameter distributions to the posteriors produced by PMCMC. Because the PMCMC algorithm requires the use of a particle filter, which is computationally infeasible for the multivariate Skellam SV model, this analysis is conducted in a univariate setting. First, consider the price changes for WMT. The computation time of PMCMC is substantial: 41 hours compared to 145 seconds of Gaussian VB and 87 seconds of Efficient VB. 

\begin{figure}[tb!]
\caption{Posterior parameter distributions for univariate Skellam stochastic volatility model}
\centering
\includegraphics*[width=\textwidth]{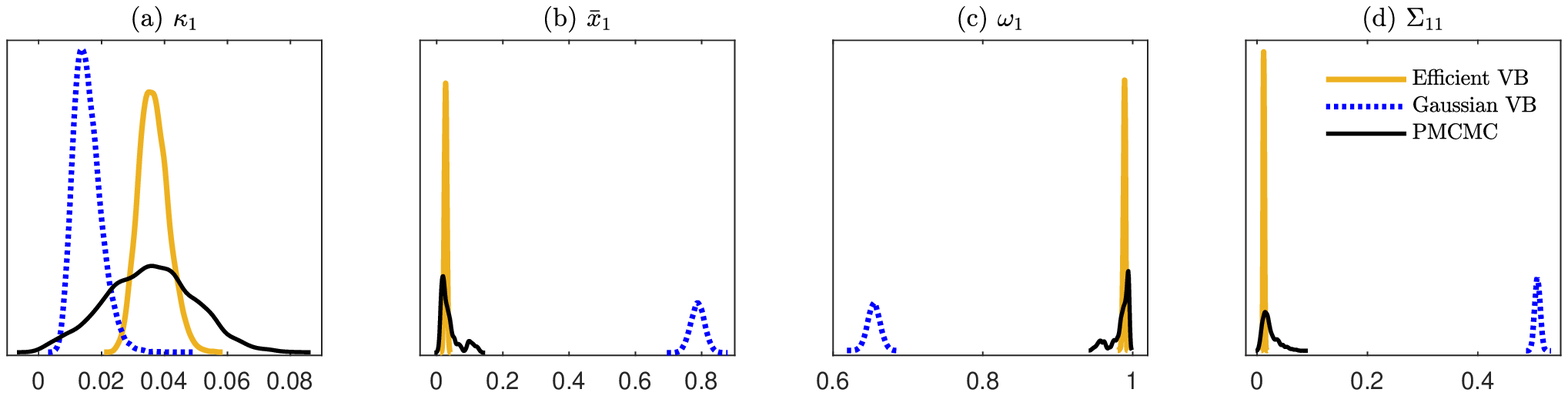}\\
\includegraphics*[width=\textwidth]{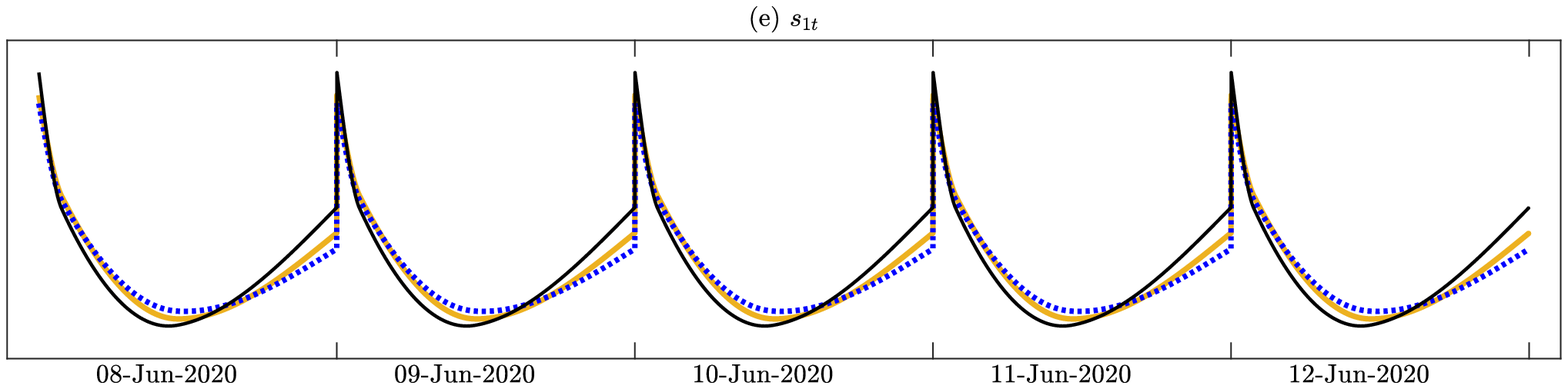}
\begin{flushleft}
This figure shows the posterior parameter distributions in the univariate Skellam stochastic volatility model for asset WMT, for Efficient VB, Gaussian VB, and PMCMC. These are indicated by the solid yellow, dotted blue, and solid black line, respectively. Panel (a)-(d) show the posterior distributions for respectively $\kappa_1$, $\bar{x}_1$, $\omega_{1}$, $\Sigma_{11}$, and Panel (e) the posterior mean for $\{s_{1t}\}_{t=1}^T$.
\end{flushleft}
\label{fig:skellam_uni}
\end{figure}

Figure~\ref{fig:skellam_uni} shows the posterior parameter distributions for $\kappa_1$, $\bar{x}_1$, $\omega_{1}$, and $\Sigma_{11}$, and the posterior mean for $\{s_{1t}\}_{t=1}^T$ in the Skellam stochastic volatility model. The posterior of Efficient VB is close to the location of the posterior of PMCMC for all four parameters, but underestimates the posterior variances. The posterior mean of Efficient VB is close to the posterior mean of PMCMC for the seasonality component across all intra-day periods. The location of the posterior distribution of Gaussian VB is less accurate for the four parameters and the seasonality.


We find similar results for the univariate Skellam stochastic volatility models for the other stocks. One way to summarize the improvement in accuracy of Efficient VB over Gaussian VB, is by comparing the ELBO values for the augmented posterior of both methods. The ELBO values averaged over the final 100 VB iterations divided by one thousand for the stocks WMT, KO, JPM, CAT equal -21.033,	-14.852,	-24.754 and	-25.002 for Gaussian VB and -20.737,	-14.508,	-24.477 and 	-24.745  for Efficient VB, respectively. These numbers summarize the accuracy of the variational approximations to the augmented posterior, with larger numbers being preferred. Efficient VB produces larger ELBO’s for all stocks.

\subsubsection{Multivariate case}
Next, we show that our method can be implemented efficiently to the multivariate Skellam stochastic volatility model with four stocks. The computation time is 18.2 minutes. This is substantially faster than MCMC with only one stock, which takes 41 hours as discussed in the previous section. Here we discuss the posterior parameter distributions and the posterior time-varying conditional volatilities of Efficient VB in the multivariate case.

First, Table~\ref{tab:skellam_multi} shows the posterior means for the multivariate Skellam stochastic volatility model estimated by Efficient VB. The $\kappa_i$ can be interpreted as the percentage of price changes that are set to zero as a result of a temporary market freezing. CAT has a relatively large amount of zero-inflation compared to almost no zero-inflation in JPM. CAT and JPM are also the stocks with the highest volatility level according to the posterior means of $\bar{x}_i$. The persistence ($\omega_i$) and variance ($\Sigma_{ii}$) in the volatilities are similar across assets. 
The posterior means of the correlations show the strong dependence between the volatility of the assets, which cannot be captured with univariate models. The posterior mean of the correlations vary between 0.579 (WMT and JPM) and 0.712 (CAT and JPM).

\begin{table}[tb!]
  \centering
  \caption{Posterior means for the multivariate Skellam stochastic volatility model}
  \begin{threeparttable}
  \begin{tabular}{cc} 
    \begin{tabular}{lrrrrr}
    \toprule \toprule
    Asset & \multicolumn{1}{l}{$i$} & \multicolumn{1}{l}{$\kappa_i$} & \multicolumn{1}{l}{$\bar{x}_i$} & \multicolumn{1}{l}{$\omega_i$} & \multicolumn{1}{l}{$\Sigma_{ii}$} \\
    \midrule
    WMT   & 1     & 0.018 & 0.240 & 0.894 & 0.148 \\
    KO    & 2     & 0.055 & 0.076 & 0.904 & 0.135 \\
    JPM   & 3     & 0.002 & 0.412 & 0.872 & 0.138 \\
    CAT   & 4     & 0.117 & 0.449 & 0.873 & 0.167 \\
          \bottomrule \bottomrule
    \end{tabular}
    &
    \begin{tabular}{lrrr}
    \toprule \toprule
     \multicolumn{4}{c}{Correlations implied by $\Sigma$}\\
    Asset &  \multicolumn{1}{l}{KO} & \multicolumn{1}{l}{JPM} & \multicolumn{1}{l}{CAT} \\
    \midrule
    WMT   &  0.654 & 0.579 & 0.609 \\
    KO    &        & 0.636 & 0.644 \\
    JPM   &        &       & 0.712 \\
          \bottomrule \bottomrule
    \end{tabular}%
    \end{tabular} 
\begin{tablenotes}
\footnotesize
\item This table shows the posterior means for the multivariate Skellam stochastic volatility model with assets WMT, KO, JPM, and CAT estimated by Efficient VB. The columns show the posterior means for $\kappa_i$, $\bar{x}_i$, $\omega_{i}$, $\Sigma_{ii}$, and the correlations implied by $\Sigma$. 
\end{tablenotes}
\end{threeparttable}
  \label{tab:skellam_multi}%
\end{table}%

\begin{figure}[tb!]
\caption{Posterior mean of the time-varying conditional volatilities}
\centering
\includegraphics*[width=\textwidth]{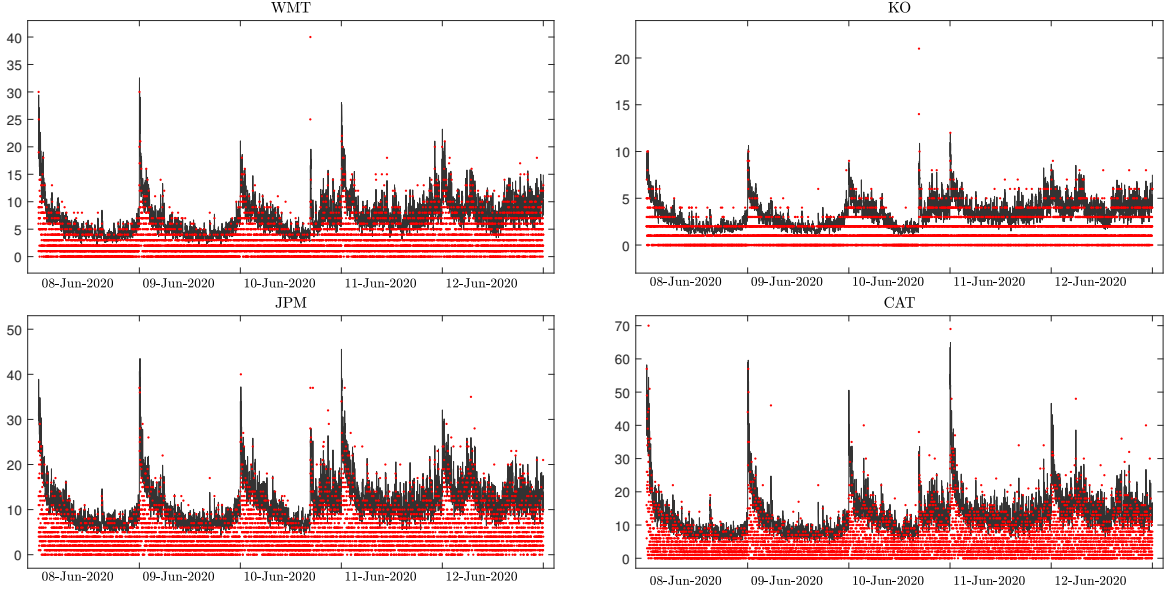}
\begin{flushleft}
This figure shows the posterior mean of sd$(y_{i,t}|x_{i,t},\theta)=\sqrt{(1-\kappa_i)\sigma_{i,t}^2}$  estimated with Efficient VB and rescaled by a factor 2, together with the absolute value of the price changes. These are indicated by the solid  black line and red dots, respectively. Panel (a)-(d) show the trading days between 8-12 June 2020 for the assets WMT, KO, JPM, and CAT, respectively.
\end{flushleft}
\label{fig:skellam_multi}
\end{figure}

Second, Figure~\ref{fig:skellam_multi} shows the posterior mean of the time-varying conditional volatilities, together with the absolute value of the price changes. The conditional volatilities exhibit the correct volatility level and follow the daily seasonality in the data, with a spike in the volatility at the start of each trading day. Additionally, conditional volatilities peak at times of outliers in the observed price changes and show persistence over time. We conclude that the intradaily patterns in the magnitude of the price variations are captured in the conditional volatilities. 

\section{The time-varying parameter VAR stochastic volatility model}\label{sec:application}
This second empirical application of Efficient VB fits a time-varying parameter vector autoregression with a stochastic volatility model (TVP-VAR-SV) to eight macroeconomic variables. This application shows that our approach is also fast and accurate in a model with a nonlinear measurement equation and a high-dimensional state vector. In contrast to the empirical application in Section~\ref{sec:skellam}, Hybrid VB is also feasible in this model, which allows us to show that estimation by Efficient VB is faster with a negligible loss in accuracy.

The data contains 150 quarterly observations from 1980:Q3 to 2017:Q4 on eight macroeconomic variables. The FRED mnemonics for these variables are GDPC1, PCECC96, FPIx, CE16OV, CES0600000007, GDPCTPI, CES0600000008, and FEDFUNDS. We fit a TVP-VAR-SV with a lag length of 2. The data set is described in detail by \citet{huber2021inducing}. 

\subsection{The model}
The VAR representation of the model is 
\begin{align}\label{eq:tvpvarsv}
\yvec_t &=  \betavec_{0,t}+\sum_{s=1}^p B_{s,t}\yvec_{t-s}+L_t^{-1}\epsilonvec_t,  \quad\epsilonvec_t \sim N(\bm{0},H_t),\nonumber\\
\betavec_t &= \betavec_{t-1} + \wvec_t,  \qquad\qquad\qquad\qquad\wvec_t \sim N(0,V), \nonumber  \\
h_{i,t} &=  \bar{h}_i+\rho_i(h_{i,t-1}-\bar{h}_i)+e_{i,t},  \quad e_{i,t} \sim N(0,\sigma_i^2),\quad \mbox{ for }
i=1,\ldots,N, 
\end{align}
where $\yvec_t=(y_{1,t},y_{2,t},\ldots,y_{N,t})^\top$ represents the $N$ macroeconomic variables at time $t$, $L_t^{-1}$ is a lower triangular matrix with unit-valued diagonal elements and lower-diagonal elements denoted as $\bm{l}_t$, $\betavec_{0,t}$ is the intercept vector, $B_{1,t},\ldots,B_{p,t}$ 
are $(N\times N)$ autoregressive coefficient matrices, and $H_t=\mbox{diag}(e^{h_{1,t}},\ldots,e^{h_{N,t}})$ is a diagonal matrix.
The $K=(pN^2+N+N(N-1)/2)$ time-varying coefficients are collected in the $K$-dimensional vector $\betavec_t^\top\equiv
(\betavec_{0,t}^\top,\mbox{vec}(B_{1,t})^\top,\ldots,\mbox{vec}(B_{p,t})^\top,\bm{l}_t^\top)$ and $V=\mbox{diag}(v_1,\ldots,v_K)$ is a diagonal matrix. 
The logarithms of the volatilities
$h_{i,1},\ldots,h_{i,T}$ follow a stationary
first order autoregression with mean $\bar{h}_i$ and autoregressive parameter $|\rho_i|<1$.
We use a horseshoe prior to regularize the time-varying parameters, as proposed by \citep{huber2021inducing}.

Estimation of the joint model in \eqref{eq:tvpvarsv} is difficult, and therefore it is common to transform the VAR model to $N$ unrelated regressions \citep{carriero2019large,kastner2020sparse}. Moreover, horseshoe priors are known to result in posterior densities that are difficult to approximate \citep{ghosh2019model}, which can be solved by adopting the re-parametrization proposed by \citet{ingraham2017variational}. Appendix~\ref{A:var} shows that after these two transformations, \eqref{eq:tvpvarsv} can be represented by $i=1,\dots,N$ state space models:
\begin{align}\label{eq:ssm_tvpvarsv}
p(y_{i,t}|\mathbf{x}_{i,t},\bm\theta_i)& = \phi_1(y_{i,t};({\bm{z}}_{i,t}^\top,{\bm{z}}_{i,t}^\top\text{diag}(\tilde{\bm{\eta}}_{i,t}))\alphavec_i,e^{h_{i,t}}),\nonumber\\ 
p(\mathbf{x}_{i,t}|\mathbf{x}_{i,t-1},\bm{\theta}_i)& = \phi_{(Np+i+1)}(\mathbf{x}_{i,t};\bar{\mathbf{x}}_i+A_{1,i}\mathbf{x}_{i,t-1},A_{2,i}^2),\,
\end{align}
where $\mathbf{x}_{i,t} = (\tilde{\bm{\eta}}_{i,t}^\top,h_{i,t})^\top$ is the $(Np+i+1)-$dimensional state vector, with $\tilde{\bm{\eta}}_{i,t}^\top$ a function of the coefficient vector $\bm\beta_t$. The parameter vector for equation $i$ is defined as $\bm{\theta}_i=\left(\bm{\tau}_i^\top,\bm{\chi}_i^\top,\xi_i,\bar{h}_i,\rho_i,\sigma_i^2\right)^\top$, with $\bm{\alpha}_i = \sqrt{\xi_i}(\bm{\tau}_i\circ\sqrt{\bm{\chi}_i})$ and $J_i$-dimensional parameter vectors $\tauvec_i = (\tau_{i,1},\dots,\tau_{i,J_i})^\top$ and $\sqrt{\bm{\chi}_i} = \left(\chi_{i,1}^{1/2},\dots,\chi_{i,J_i}^{1/2}\right)^\top$ with $J_i=2(pN+i)$, and scalar parameter $\xi_i$.
The ($p+N-1$)-dimensional vector ${\bm{z}}_{i,t}=\left(\bm{y}_{t-1}^\top,\dots,\bm{y}_{t-p}^\top,1,-\bm{y}_{1:i-1,t}^\top\right)^\top$ with $\bm{y}_{1:i-1,t} = \left(y_{1,t},\dots,y_{i-1,t}\right)^\top$, represents the covariates in the measurement density. The parameters in the state density $\bar{\mathbf{x}}_i$, $A_{1,i}$ and $A_{2,i}$ are functions of $\rho_i$, $\sigma_i$ and $\bar{h}_i$. Here, $A_{2,i}^2$ denotes the operation of squaring each of the elements in $A_{2,i}.$

Since the state vector $\xvec_{i,t}$ is high-dimensional and enters the measurement equation non-linearly via $h_{i,t}$, the states cannot be analytically integrated out of the likelihood function. Hence, we consider the augmented  posterior distribution. Let $\yvec_{(i)} \equiv (y_{i,1},\ldots,y_{i,T})^\top$ be the observations on the $i$th macroeconomic variable, $\yvec_{(\backslash i)}$ be the observations on the other $N-1$ macroeconomic variables, and $\xvec_{(i)}\equiv (\xvec_{i,1}^\top,\ldots,\xvec_{i,T}^\top)^\top$ the latent states in the $i$th equation. The augmented posterior is
\begin{align}\label{eq:augpost}
p(\thetavec_i,\xvec_{(i)}|\yvec) \propto p(\yvec_{(i)}|\xvec_{(i)},\yvec_{(\backslash i)})p(\xvec_{(i)}|\thetavec_i)p(\thetavec_i)
 =
\prod_{t=1}^{T}\left\{\phi_{1}\left(y_{i,t};({\bm{z}}_{i,t}^\top,{\bm{z}}_{i,t}^\top\text{diag}(\tilde{\bm{\eta}}_{i,t}))
\alphavec_i
,e^{h_{i,t}}\right)\right\}\times\\
\phi_{(Np+i+1)}\left( \xvec_{i,1}; \bar{\xvec}_{i,1},V_{i,1}\right)  
\prod_{t=2}^T \left\{\phi_{Np+i+1}\left( \xvec_{i,t}; \bar{\mathbf{x}}_i+A_{1,i}\mathbf{x}_{i,t-1},A_{2,i}^2\right)\right\}
p(\thetavec_i),
\end{align}
where $\bar{\xvec}_{i,1} = (\bm{0}_{Np+i}^\top,\bar{h}_i)^\top$ and $V_{i,1} = \text{diag}((\bm{1}_{Np+i}^\top,\frac{\sigma_i^2}{1-\rho_i^2}))$.
The prior density for $\bm\theta_i$ is specified as $p(\thetavec_i)=p(\xi_{i}|\kappa_{i})p(\kappa_i)p(\bar{h}_i)p(\rho_i)p(\sigma_i^2)\prod_{j=1}^{J_i}p(\tau_{i,j})p(\chi_{i,j}|\nu_{i,j})p(\nu_{i,j})$, with $p(\xi_{i}|\kappa_{i})=$Inverse-Gamma(0.5,$\kappa_i^{-1})$, $p(\kappa_i)=$Inverse-Gamma(0.5,1), $p(\bar{h}_i)=N(0,100)$, $p((\rho_i+1)/2)=$Beta(25,5), $p(\sigma_i^2)$=Gamma(0.5,0.5), $p(\tau_{i,j})=N(0,1)$, $p(\chi_{i,j}|\nu_{i,j})=$Inverse-Gamma(0.5,$\nu_{i,j}^{-1})$, and $p(\nu_{i,j})=$Inverse-Gamma(0.5,1).
The MCMC sampler from the augmented posterior in \eqref{eq:augpost} is discussed by \citet{huber2021inducing}.

\subsection{Variational approximations}


Each separate augmented posterior of the model admits an approximation as proposed in \eqref{eq:qxy}. 
%
Since the state transition is a multivariate Gaussian, the sufficient summary vector is $\bm T( \mathbf{x}_t)=(\mathbf{x}_{i,t}^\top,\text{vec}(\mathbf{x}_{i,t}\mathbf{x}_{i,t}^\top))^\top$. We define the vector of kernel parameters to be $\bm a_{i,t} = (\mathbf{b}_{i,t},\text{vec}(C_{i,t}))^\top$ with $\bm{b}_{i,t}$ an $(Np+i+1)$-dimensional vector and $C_{i,t} = \text{diag}(\bm{c}_{i,t})$ specified as a diagonal matrix for computational efficiency, where $\bm{c}_{i,t}$ is an  $(Np+i+1)$-dimensional vector. 
%

The approximation to the states $q(\mathbf{x}_{(i)}|\mathbf{y}) = \prod_{t=1}^Tq(\mathbf{x}_{i,t}|\mathbf{x}_{i,t-1},\mathbf{y},\bm\varphi_i)$ is a product of Gaussian densities such that $q(\mathbf{x}_{i,t}|\mathbf{x}_{i,t-1},\mathbf{y},\bm\varphi_i) = \phi_{(Np+i+1)}(\mathbf{x}_{i,t};\bm\mu_{i,t},\Sigma_{i,t})$ with 
 $\Sigma_{i,t} = \left(A_{2,i}^{-2}-2C_{i,t}\right)^{-1}$, and $\bm{\mu}_{i,t} = \Sigma_{i,t} \left(\bm{b}_{i,t}+A_{2,i}^{-2}\left(\bar{\mathbf{x}}_i+A_{i,1}\mathbf{x}_{i,t-1}\right)\right)$.
 The integration constant of the transition kernel equals
\begin{align}
  \chi(\mathbf{x}_{i,t-1}|\bm a_{i,t},\bm\varphi_i)
  &=  \exp\left[ \frac{1}{2}\log \frac{|\Sigma_{i,t}|}{|A_{2,i}^2|}+\frac{1}{2}\bm{\mu}_{i,t}^\top\Sigma_{i,t}^{-1}\bm{\mu}_{i,t}-\frac{1}{2}\left(\bar{\mathbf{x}}_i+A_{1,i}\mathbf{x}_{i,t-1}\right)^\top A_{2,i}^{-2}\left(\bar{\mathbf{x}}_i+A_{1,i}\mathbf{x}_{i,t-1}\right) \right]. \notag
\end{align}
While not made explicit in the notation, the parameters $\bar{\mathbf{x}}_i$, $A_{1,i}$ and $A_{2,i}$ are determined by the proxy parameter vector $\bm\varphi_i$. 
Since we assume $C_{i,t}$ to be diagonal, only the kernel parameters $\bm{b}_{i,t}$ and $\bm{c}_{i,t}$ have to be calibrated. Hence, $\bm{\gamma}_{i,t}^\top\bm{T}(\mathbf{x}_{i,t}^{[s]})$ boils down to $\bm{\tilde{\gamma}}_{i,t}^\top(\mathbf{x}_{i,t}^\top,(\mathbf{x}_{i,t}^2)^\top )^\top$ in Algorithm~\ref{alg:vi2} in Appendix~\ref{app:calibration}.

In addition to our method, we also consider Gaussian and Hybrid variational approximations. All three VB methods use a Gaussian approximation with a factor covariance matrix for $\bm{\theta}_i$ and set the number of factors to one. The gradients for our variational approximation together with  implementation details of the benchmark methods are provided in \cite{loaiza2022fast}. 
MCMC is implemented using a burn-in sample of size 15,000 and inference sample of size 15,000.
We run all three VB algorithms for a total of 10,000 iterations. Appendix~\ref{A:application} shows that this number of iterations is suffices to achieve convergence.

\subsection{Results}

 The computation time of Efficient VB is 2.170 minutes. This is faster than Gaussian VB and Hybrid VB, which takes 2.452 and 8.791 minutes, respectively. Since MCMC takes 26.390 minutes, Efficient VB uses less than 9\% of the time required for MCMC. 


%

\subsubsection{Posterior distribution of the states}
The TVP-VAR-SV model in \eqref{eq:tvpvarsv} contains a total of 172 states at each of the 150 time periods. To illustrate the posterior estimates for the state vectors, we consider the posterior mean of one of the time-varying VAR coefficients and of one of the time-varying volatilities.

\begin{figure}[tb!]
\caption{Posterior mean of a time-varying VAR coefficient}
\centering
\includegraphics*[width=\textwidth]{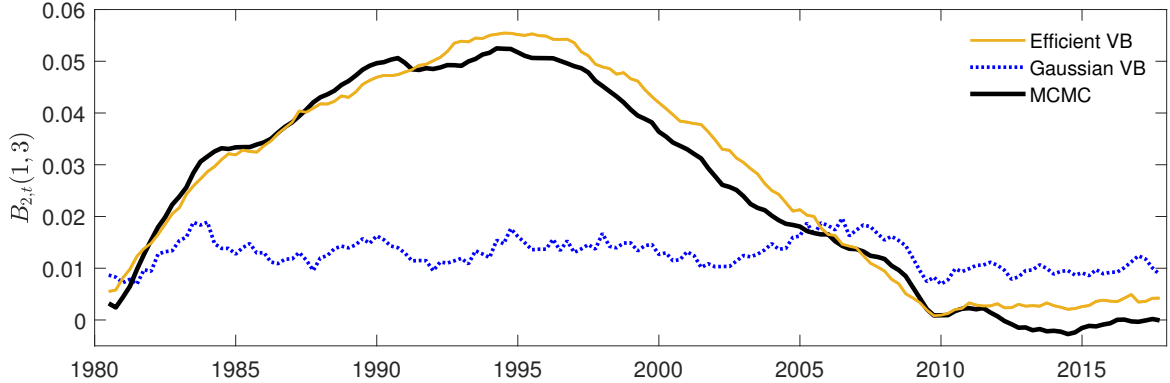}\\[3mm]
\begin{flushleft}
This figure shows the posterior mean of $B_{2,t}(1,3)$ in \eqref{eq:tvpvarsv} across time $t$, for Efficient VB, Gaussian VB, and MCMC, indicated by the solid yellow, dotted blue, and solid black line, respectively.
\end{flushleft}
\label{fig:Coefstates}
\end{figure}

First, Figure~\ref{fig:Coefstates} shows the posterior mean of one of the time-varying VAR coefficients in \eqref{eq:tvpvarsv}, $B_{2,t}(1,3)$, across time $t$, for Efficient VB, Gaussian VB, and MCMC. Remember that Hybrid VB uses the exact conditional density of the states in its variational approximation, and hence its posterior mean for the states is very similar to MCMC and not included in the figure. Many of the time-varying VAR coefficients are regularized to zero with all methods. Therefore we illustrate the posterior state distributions by the posterior mean of $B_{2,t}(1,3)$, a coefficient that actually has time dynamics and differences in these dynamics across the different methods. 

The posterior mean for Efficient VB is similar to the posterior mean for MCMC over the whole sample period. Both show substantial variation over time, with a posterior mean close to zero in 1980, increasingly positive between 1980 and 1995, decreasing between 1995 and 2010, and close to zero again between 2010 and 2017. The posterior mean for Gaussian VB follows a different time path with a small amount of variation and close to zero across the whole sample. 

\begin{figure}[tb!]
\caption{Posterior mean of a time-varying volatility}
\centering
\includegraphics*[width=\textwidth]{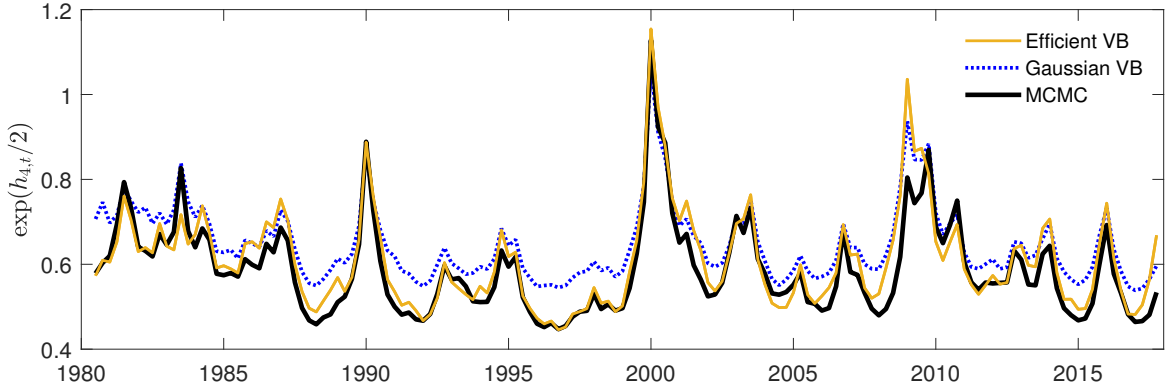}\\[3mm]
\begin{flushleft}
This figure shows the posterior mean of $\exp(h_{4,t}/2)$ in \eqref{eq:tvpvarsv} across time $t$, for Efficient VB, Gaussian VB, and MCMC, indicated by the solid yellow, dotted blue, and solid black line, respectively.
\end{flushleft}
\label{fig:Volstates}
\end{figure}

Second, Figure~\ref{fig:Volstates} shows the posterior mean of one of the time-varying volatilities in \eqref{eq:tvpvarsv}, $\exp(h_{4,t}/2)$, across time $t$. The results are similar to the posterior means for the time-varying coefficients. Efficient VB more accurately approximates the posterior mean for MCMC than Gaussian VB. Similar to the findings for Figure~\ref{fig:SVstates} in the numerical experiment, Gaussian VB seems to overestimate states compared to the posterior means for MCMC. Figure \ref{fig:VolstatesAll} in Appendix~\ref{A:application} shows that a similar conclusion can be drawn when looking at the posterior mean for the time-varying volatilities of the remaining equations. 

\subsubsection{Posterior distribution of the parameters}
\begin{figure}[tb!]
\caption{Posterior distributions for parameters in the stochastic volatility component}
\centering
\includegraphics*[scale=0.7]{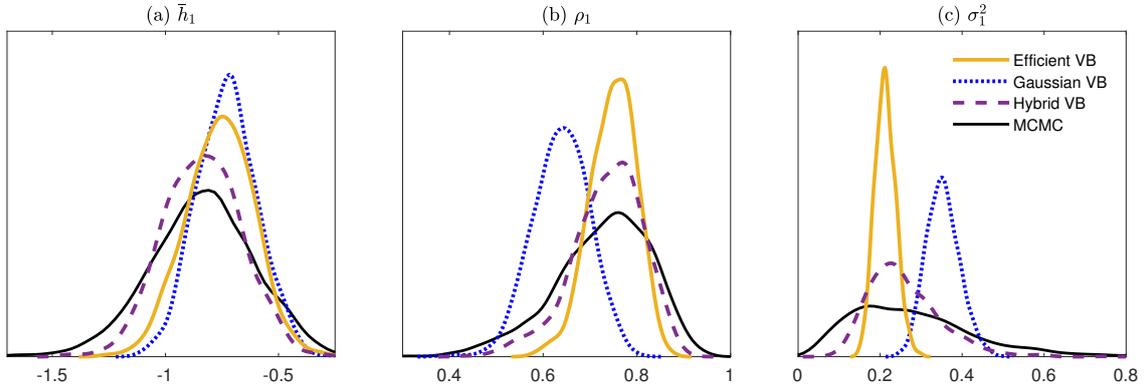}\\[3mm]
\begin{flushleft}
This figure shows the posterior parameter distributions in the stochastic volatility component for variable 1 in \eqref{eq:tvpvarsv}, for Efficient VB, Gaussian VB, Hybrid VB, and MCMC. These are indicated by the solid yellow, dotted blue, dashed purple, and solid black line, respectively. Panel (a) shows the posterior distribution for $\bar{h}_1$, Panel (b) for $\rho_1$, and Panel (c) for $\sigma_1^2$. 
\end{flushleft}
\label{fig:Param}
\end{figure}

To assess the accuracy of the posterior distribution for the parameters, we focus on the parameters in the stochastic volatility model for Real Gross Domestic
Product (GDPC1), which is the first variable in \eqref{eq:tvpvarsv}. Figure~\ref{fig:Param} shows the posterior parameter distributions for $\bar{h}_1$, $\rho_1$, and $\sigma_1^2$. The posterior of Efficient VB is close to the location of  the posterior of MCMC for all three parameters, but underestimates the posterior variances. Although slightly more accurate, we find the same for Hybrid VB. The location of the posterior distribution of Gaussian VB is less accurate for $\rho_1$ and $\sigma_1^2$. 

Table~\ref{tab:elbo} shows the ELBO values for the augmented posterior for both Efficient VB and Gaussian VB, for each equation in \eqref{eq:ssm_tvpvarsv} averaged over the final 100 VB iterations divided by one thousand. These numbers summarize the accuracy of the variational approximations to the augmented posterior, with larger numbers being preferred. We find that Efficient VB produces larger ELBO's for all equations.

\begin{table}[tb!]
  \centering
  \caption{ELBO values for each state space model in \eqref{eq:ssm_tvpvarsv}}
  \begin{threeparttable}
    \begin{tabular}{lrrrrrrrr}
    \toprule \toprule
    Equation & 1     & 2     & 3     & 4     & 5     & 6     & 7     & 8 \\ \midrule
    Efficient VB & -0.140 & -0.136 & -0.109 & -0.165 & -0.179 & -0.157 & -0.160 & -0.239 \\
    Gaussian VB & -0.960 & -1.005 & -1.031 & -1.107 & -1.181 & -1.262 & -1.296 & -1.244 \\
          \bottomrule \bottomrule
    \end{tabular}%
\begin{tablenotes}
\footnotesize
\item This table shows the ELBO values for each equation $i$ in \eqref{eq:ssm_tvpvarsv} averaged over the final 100 VB iterations divided by one thousand, for Efficient VB and Gaussian VB. Note that the ELBO cannot be calculated for Hybrid VB.
\end{tablenotes}
\end{threeparttable}
  \label{tab:elbo}
\end{table}

The results are similar to the findings for Figure~\ref{fig:SVsmall} in the numerical experiment, with a smaller sample and a more complex model. If the empirical results are consistent with the numerical analyses in Figure~\ref{fig:SVlarge}, the bias in Gaussian VB is expected to increase as the sample grows, while Efficient VB is not expected to lose accuracy. 

\section{Conclusion}\label{sec:conclusion}
This paper proposes a variational Bayes method for state space models, that uses a new variational approximation to the states. This approximation conditions on the observed data, which results in an accurate approximation to the posterior distribution of both the states and the model parameters. Since the approximation is calibrated in a computationally efficient way, the method is fast and scalable to a large number of states and a large number of observations. The combination of accuracy and speed of the variational approximation is illustrated in numerical experiments with a simple stochastic volatility model and empirical applications to a novel multivariate Skellam stochastic volatility model for high-frequency tick-by-tick discrete price changes and a modern macroeconomic time-varying parameter vector autoregression with stochastic volatility. 

The proposed efficient variational Bayes method is applicable to a wide range of state space models, including  models for which accurate estimation is computationally infeasible using existing methods. First, our method can be applied to many models with nonlinear or non-Gaussian measurement equations and/or high-dimensional state vectors. For instance, two potential additional applications of the approach are dynamic stochastic copula models \citep{hafner2012dynamic}, and multivariate stochastic volatility models with realised volatility \citep{yamauchi2020multivariate}. 

Second, the method can be applied to models with any transition density that is a member of the exponential family of distributions. This opens up the possibility of applying state space modelling with a non-Gaussian transition density to, for instance, realised covariance matrices of asset returns.
The uptake of state space models in this literature has been limited, since high-dimensional state vectors with nonlinear restrictions are generally required \citep{gribisch2022modeling}. 
%
One future extension of our method is the accommodation of transition equations that are not a member of the exponential family, as is the case with the Heston model \citep{eraker2004stock}.

\bibliographystyle{apalike} 
\bibliography{vissm}

\begin{thebibliography}{}

\bibitem[Andrieu et~al., 2010]{andrieu2010particle}
Andrieu, C., Doucet, A., and Holenstein, R. (2010).
\newblock Particle {M}arkov chain {M}onte {M}arlo methods.
\newblock {\em Journal of the Royal Statistical Society: Series B (Statistical
  Methodology)}, 72(3):269--342.

\bibitem[Archer et~al., 2015]{archer2015black}
Archer, E., Park, I.~M., Buesing, L., Cunningham, J., and Paninski, L. (2015).
\newblock Black box variational inference for state space models.
\newblock {\em arXiv preprint arXiv:1511.07367}.

\bibitem[Barndorff-Nielsen et~al., 2009]{barndorff2009realized}
Barndorff-Nielsen, O.~E., Hansen, P.~R., Lunde, A., and Shephard, N. (2009).
\newblock Realized kernels in practice: Trades and quotes.

\bibitem[Carriero et~al., 2019]{carriero2019large}
Carriero, A., Clark, T.~E., and Marcellino, M. (2019).
\newblock Large {B}ayesian vector autoregressions with stochastic volatility
  and non-conjugate priors.
\newblock {\em Journal of Econometrics}, 212(1):137--154.

\bibitem[Carter and Kohn, 1994]{carter1994gibbs}
Carter, C.~K. and Kohn, R. (1994).
\newblock On {G}ibbs sampling for state space models.
\newblock {\em Biometrika}, 81(3):541--553.

\bibitem[Catania et~al., 2022]{catania2022dynamic}
Catania, L., Di~Mari, R., and Santucci~de Magistris, P. (2022).
\newblock Dynamic discrete mixtures for high-frequency prices.
\newblock {\em Journal of Business \& Economic Statistics}, 40(2):559--577.

\bibitem[Chan, 2022]{chan2022large}
Chan, J.~C. (2022).
\newblock Large hybrid time-varying parameter {VARs}.
\newblock {\em Journal of Business \& Economic Statistics},
  (just-accepted):1--34.

\bibitem[Chan and Jeliazkov, 2009]{chan2009efficient}
Chan, J.~C. and Jeliazkov, I. (2009).
\newblock Efficient simulation and integrated likelihood estimation in state
  space models.
\newblock {\em International Journal of Mathematical Modelling and Numerical
  Optimisation}, 1(1-2):101--120.

\bibitem[Chan and Yu, 2022]{chan2022fast}
Chan, J.~C. and Yu, X. (2022).
\newblock Fast and accurate variational inference for large {B}ayesian {VARs}
  with stochastic volatility.
\newblock {\em Journal of Economic Dynamics and Control}, 143:104505.

\bibitem[Chopin et~al., 2020]{chopin2020introduction}
Chopin, N., Papaspiliopoulos, O., et~al. (2020).
\newblock {\em An introduction to sequential Monte Carlo}.
\newblock Springer.

\bibitem[Clark and Ravazzolo, 2015]{clark2015macroeconomic}
Clark, T.~E. and Ravazzolo, F. (2015).
\newblock Macroeconomic forecasting performance under alternative
  specifications of time-varying volatility.
\newblock {\em Journal of Applied Econometrics}, 30(4):551--575.

\bibitem[Cross et~al., 2021]{cross2021macroeconomic}
Cross, J.~L., Hou, C., Koop, G., and Poon, A. (2021).
\newblock Macroeconomic forecasting with large stochastic volatility in mean
  {VARs}.

\bibitem[Doucet et~al., 2015]{doucet2015efficient}
Doucet, A., Pitt, M.~K., Deligiannidis, G., and Kohn, R. (2015).
\newblock Efficient implementation of {M}arkov chain {M}onte {C}arlo when using
  an unbiased likelihood estimator.
\newblock {\em Biometrika}, 102(2):295--313.

\bibitem[Eraker, 2004]{eraker2004stock}
Eraker, B. (2004).
\newblock Do stock prices and volatility jump? {R}econciling evidence from spot
  and option prices.
\newblock {\em The Journal of finance}, 59(3):1367--1403.

\bibitem[Frazier et~al., 2022]{frazier2022variational}
Frazier, D.~T., Loaiza-Maya, R., and Martin, G.~M. (2022).
\newblock Variational {B}ayes in state space models: Inferential and predictive
  accuracy.
\newblock Technical report, Monash University, Department of Econometrics and
  Business Statistics.

\bibitem[Gefang et~al., 2022]{gefang2019variational}
Gefang, D., Koop, G., and Poon, A. (2022).
\newblock Forecasting using variational {B}ayesian inference in large vector
  autoregressions with hierarchical shrinkage.
\newblock {\em International Journal of Forecasting}, pages 1--18.

\bibitem[Ghosh et~al., 2019]{ghosh2019model}
Ghosh, S., Yao, J., and Doshi-Velez, F. (2019).
\newblock Model selection in {B}ayesian neural networks via horseshoe priors.
\newblock {\em J. Mach. Learn. Res.}, 20(182):1--46.

\bibitem[Greenberg, 2012]{greenberg2012introduction}
Greenberg, E. (2012).
\newblock {\em Introduction to Bayesian econometrics}.
\newblock Cambridge University Press.

\bibitem[Gribisch and Hartkopf, 2022]{gribisch2022modeling}
Gribisch, B. and Hartkopf, J.~P. (2022).
\newblock Modeling realized covariance measures with heterogeneous liquidity: a
  generalized matrix-variate {W}ishart state-space model.
\newblock {\em Journal of Econometrics}.

\bibitem[Hafner and Manner, 2012]{hafner2012dynamic}
Hafner, C.~M. and Manner, H. (2012).
\newblock Dynamic stochastic copula models: Estimation, inference and
  applications.
\newblock {\em Journal of Applied Econometrics}, 27(2):269--295.

\bibitem[Harvey and Koopman, 1993]{harvey1993forecasting}
Harvey, A. and Koopman, S.~J. (1993).
\newblock Forecasting hourly electricity demand using time-varying splines.
\newblock {\em Journal of the American Statistical Association},
  88(424):1228--1236.

\bibitem[Huber et~al., 2021]{huber2021inducing}
Huber, F., Koop, G., and Onorante, L. (2021).
\newblock Inducing sparsity and shrinkage in time-varying parameter models.
\newblock {\em Journal of Business \& Economic Statistics}, 39(3):669--683.

\bibitem[Ingraham and Marks, 2017]{ingraham2017variational}
Ingraham, J. and Marks, D. (2017).
\newblock Variational inference for sparse and undirected models.
\newblock In {\em International Conference on Machine Learning}, pages
  1607--1616. PMLR.

\bibitem[Kastner and Huber, 2020]{kastner2020sparse}
Kastner, G. and Huber, F. (2020).
\newblock Sparse {B}ayesian vector autoregressions in huge dimensions.
\newblock {\em Journal of Forecasting}, 39(7):1142--1165.

\bibitem[Kim et~al., 1998]{kim1998stochastic}
Kim, S., Shephard, N., and Chib, S. (1998).
\newblock Stochastic volatility: likelihood inference and comparison with
  {ARCH} models.
\newblock {\em The review of economic studies}, 65(3):361--393.

\bibitem[Kingma and Welling, 2013]{kingma2013auto}
Kingma, D.~P. and Welling, M. (2013).
\newblock Auto-encoding variational {B}ayes.
\newblock {\em arXiv preprint arXiv:1312.6114}.

\bibitem[Koop and Korobilis, 2018]{koop2018variational}
Koop, G. and Korobilis, D. (2018).
\newblock Variational {B}ayes inference in high-dimensional time-varying
  parameter models.

\bibitem[Koopman et~al., 2017]{koopman2017intraday}
Koopman, S.~J., Lit, R., and Lucas, A. (2017).
\newblock Intraday stochastic volatility in discrete price changes: the dynamic
  {S}kellam model.
\newblock {\em Journal of the American Statistical Association},
  112(520):1490--1503.

\bibitem[Koopman et~al., 2018]{koopman2018dynamic}
Koopman, S.~J., Lit, R., Lucas, A., and Opschoor, A. (2018).
\newblock Dynamic discrete copula models for high-frequency stock price
  changes.
\newblock {\em Journal of Applied Econometrics}, 33(7):966--985.

\bibitem[Koopman et~al., 2015]{koopman2015numerically}
Koopman, S.~J., Lucas, A., and Scharth, M. (2015).
\newblock Numerically accelerated importance sampling for nonlinear
  non-{G}aussian state-space models.
\newblock {\em Journal of Business \& Economic Statistics}, 33(1):114--127.

\bibitem[Loaiza-Maya et~al., 2022]{loaiza2022fast}
Loaiza-Maya, R., Smith, M.~S., Nott, D.~J., and Danaher, P.~J. (2022).
\newblock Fast and accurate variational inference for models with many latent
  variables.
\newblock {\em Journal of Econometrics}, 230(2):339--362.

\bibitem[Naesseth et~al., 2018]{naesseth2018variational}
Naesseth, C., Linderman, S., Ranganath, R., and Blei, D. (2018).
\newblock Variational sequential {M}onte {C}arlo.
\newblock In {\em International conference on artificial intelligence and
  statistics}, pages 968--977. PMLR.

\bibitem[Ong et~al., 2018]{ong2018gaussian}
Ong, V. M.-H., Nott, D.~J., and Smith, M.~S. (2018).
\newblock Gaussian variational approximation with a factor covariance
  structure.
\newblock {\em Journal of Computational and Graphical Statistics},
  27(3):465--478.

\bibitem[Quiroz et~al., 2022]{quiroz2018gaussian}
Quiroz, M., Nott, D.~J., and Kohn, R. (2022).
\newblock Gaussian variational approximation for high-dimensional state space
  models.
\newblock {\em Bayesian Analysis}.

\bibitem[Richard and Zhang, 2007]{richard2007efficient}
Richard, J.-F. and Zhang, W. (2007).
\newblock Efficient high-dimensional importance sampling.
\newblock {\em Journal of Econometrics}, 141(2):1385--1411.

\bibitem[Scharth and Kohn, 2016]{scharth2016particle}
Scharth, M. and Kohn, R. (2016).
\newblock Particle efficient importance sampling.
\newblock {\em Journal of Econometrics}, 190(1):133--147.

\bibitem[Shephard and Yang, 2017]{shephard2017continuous}
Shephard, N. and Yang, J.~J. (2017).
\newblock Continuous time analysis of fleeting discrete price moves.
\newblock {\em Journal of the American Statistical Association},
  112(519):1090--1106.

\bibitem[Tan and Nott, 2018]{tan2018gaussian}
Tan, L.~S. and Nott, D.~J. (2018).
\newblock Gaussian variational approximation with sparse precision matrices.
\newblock {\em Statistics and Computing}, 28(2):259--275.

\bibitem[Tran et~al., 2017]{tran2017variational}
Tran, M.-N., Nott, D.~J., and Kohn, R. (2017).
\newblock Variational {B}ayes with intractable likelihood.
\newblock {\em Journal of Computational and Graphical Statistics},
  26(4):873--882.

\bibitem[Wang and Titterington, 2004]{wang2004lack}
Wang, B. and Titterington, D. (2004).
\newblock Lack of consistency of mean field and variational {B}ayes
  approximations for state space models.
\newblock {\em Neural Processing Letters}, 20(3):151--170.

\bibitem[Yamauchi and Omori, 2020]{yamauchi2020multivariate}
Yamauchi, Y. and Omori, Y. (2020).
\newblock Multivariate stochastic volatility model with realized volatilities
  and pairwise realized correlations.
\newblock {\em Journal of Business \& Economic Statistics}, 38(4):839--855.

\bibitem[Zeiler, 2012]{zeiler2012adadelta}
Zeiler, M.~D. (2012).
\newblock Adadelta: an adaptive learning rate method.
\newblock {\em arXiv preprint arXiv:1212.5701}.

\end{thebibliography}
\clearpage
\clearpage
\appendix

\section{Calibration of $\bm a$}\label{app:calibration}
The state transition density can be written as
\begin{align}
    p(\mathbf{x}_t|\mathbf{x}_{t-1},\bm\theta) = h(\mathbf{x}_t)g(\mathbf{x}_{t-1},\bm\theta)\exp\left( \bm{\eta}(\mathbf{x}_{t-1},\bm\theta)^\top \bm{T}(\mathbf{x}_t)\right),
\end{align}
where $h(.)$, $g(.,.)$, $\bm\eta(.,.)$, and $\bm T(.)$ are known analytical functions. Note that $g(\mathbf{x}_{t-1},\bm\theta)^{-1}$ is the normalizing constant of $p(\mathbf{x}_t|\mathbf{x}_{t-1},\bm\theta)$. The transition kernel $k(\mathbf{x}_t,\mathbf{x}_{t-1}|\bm a_t,\bm\varphi)$ is given by
\begin{align}
    k(\mathbf{x}_t,\mathbf{x}_{t-1}|\bm a_t,\bm\varphi) = \exp\left(\bm{a}_t^\top \bm{T}(\mathbf{x}_t)\right)h(\mathbf{x}_t)g(\mathbf{x}_{t-1},\bm\theta)\exp\left( \bm{\eta}(\mathbf{x}_{t-1},\bm\theta)^\top \bm{T}(\mathbf{x}_t)\right).
\end{align}
We have that
\begin{align}
    q(\mathbf{x}_t|\mathbf{x}_{t-1},\mathbf{y},\bm\varphi)=\frac{k(\mathbf{x}_t,\mathbf{x}_{t-1}|\bm{a}_t,\bm\varphi)}{\chi(\mathbf{x}_{t-1}|\bm{a}_t,\bm\varphi)}.
\end{align}
The normalising constant of this expression can be written as 
\begin{align}
\chi(\mathbf{x}_{t-1}|\bm{a}_t,\bm\varphi) = &\int \exp\left(\bm{a}_t^\top \bm{T}(\mathbf{x}_t)\right)p(\mathbf{x}_t|\mathbf{x}_{t-1},\bm\varphi)d\mathbf{x}_t\\
=& \int \exp\left(\bm{a}_t^\top \bm{T}(\mathbf{x}_t)\right)h(\mathbf{x}_t)g(\mathbf{x}_{t-1},\bm\theta)\exp\left( \bm{\eta}(\mathbf{x}_{t-1},\bm\theta)^\top \bm{T}(\mathbf{x}_t)\right)d\mathbf{x}_t\\
=& g(\mathbf{x}_{t-1},\bm\theta)\int \exp\left((\bm{a}_t+\bm{\eta}(\mathbf{x}_{t-1},\bm\theta))^\top \bm{T}(\mathbf{x}_t)\right)h(\mathbf{x}_t)d\mathbf{x}_t\\
=&g(\mathbf{x}_{t-1},\bm\theta)\int \exp\left(\tilde{\bm{\eta}}(\bm{a}_t,\mathbf{x}_{t-1},\bm\theta)^\top \bm{T}(\mathbf{x}_t)\right)h(\mathbf{x}_t)d\mathbf{x}_t\\
=&\frac{g(\mathbf{x}_{t-1},\bm\theta)}{\tilde{g}(\bm a_t,\mathbf{x}_{t-1},\bm\theta)},
\end{align}
with $\tilde{\bm{\eta}}(\bm{a}_t,\mathbf{x}_{t-1},\bm\theta) = \bm{a}_t+\bm{\eta}(\mathbf{x}_{t-1},\bm\theta)$, $\tilde{g}(\bm a_t,\mathbf{x}_{t-1},\bm\theta)^{-1} = \int \exp\left(\tilde{\bm{\eta}}(\bm{a}_t,\mathbf{x}_{t-1},\bm\theta)^\top \bm{T}(\mathbf{x}_t)\right)h(\mathbf{x}_t)d\mathbf{x}_t$. Note that the solution to this integral depends on the value of $\tilde{\bm{\eta}}(\bm{a}_t,\mathbf{x}_{t-1},\bm\theta)$, and therefore $\bm{a}_t$. Thus, the values $\bm{a}_t$ must be constrained to the space where $q(\mathbf{x}_t|\mathbf{x}_{t-1},\mathbf{y},\bm\varphi)$ is a valid distribution function. For instance, for a multivariate normal distribution this would imply that $\bm{a}_t$ must induce a variance-covariance matrix that is positive definite.

\cite{richard2007efficient} calibrate  $\bm{a}_t$ as the solution to the optimization problem:
\begin{align*}
    \bm{a}_t &= \argmin_{\tilde{\bm a}_t\in A_t} \sum_{s=1}^S \left(-\gamma_{0,t}+\log(p (\mathbf{y}_t|\mathbf{x}_t^{[s]},\bm\varphi)p (\mathbf{x}_t^{[s]}|\mathbf{x}_{t-1}^{[s]},\bm\varphi)\chi(\mathbf{x}_{t}^{[s]}|\tilde{\bm{a}}_{t+1},\bm\varphi))-\log(k(\mathbf{x}_t^{[s]},\mathbf{x}_{t-1}^{[s]}|\tilde{\bm a}_t,\bm\varphi)) \right)^2,
\end{align*}
where $S$ state paths are drawn as $\mathbf{x}^{[s]}\sim q(\mathbf{x}|\mathbf{y})$.
For our choice of kernel function and state transition, we can show that
\begin{align}
   -\gamma_{0,t}+\log(p (\mathbf{y}_t|\mathbf{x}_t^{[s]},\bm\varphi)p (\mathbf{x}_t^{[s]}|\mathbf{x}_{t-1}^{[s]},\bm\varphi)\chi(\mathbf{x}_{t}^{[s]}|\tilde{\bm{a}}_{t+1},\bm\varphi))-\log(k(\mathbf{x}_t^{[s]},\mathbf{x}_{t-1}^{[s]}|\tilde{\bm a}_t,\bm\varphi))=\\
     -\gamma_{0,t}+\log(p (\mathbf{y}_t|\mathbf{x}_t^{[s]},\bm\varphi)\chi(\mathbf{x}_{t}^{[s]}|\tilde{\bm{a}}_{t+1},\bm\varphi)-\tilde{\bm{a}}_t^\top\bm{T}(\mathbf{x}_t^{[s]}),
\end{align}
which induces a sequence of linear regression problems. Thus, we can set $\bm a_t = \hat{\bm\gamma}_t$, where  $\hat{\bm\gamma}_t$ is the OLS coefficient estimate of the linear regression $$\log(p (\mathbf{y}_t|\mathbf{x}_t^{[s]},\bm\varphi)\chi(\mathbf{x}_{t}^{[s]}|\bm{a}_{t+1},\bm\varphi) = \gamma_{0,t}+ \bm{\gamma}_t^\top T(\mathbf{x}_t^{[s]})+\nu_{s,t}.$$
The intercept $\gamma_{0,t}$ does not play any further role in the method. Algorithm \ref{alg:vi2} summarises the implementation details.

\begin{algorithm}
    \begin{algorithmic}[1]
        \State{Choose an initial value  $\bm a$.}
         \State{Generate $S$ state paths $\mathbf{x}^{[s]}\sim q(\mathbf{x}|\mathbf{y})$.}
        \For{$t = T,\dots,1$}
        \State{Set $\tilde{{y}}_{s,t} = \log \left[p (\mathbf{y}_t|\mathbf{x}_t^{[s]},\bm\varphi)\chi(\mathbf{x}_{t}^{[s]}|\bm{a}_{t+1},\bm\varphi)\right]$  for $s = 1,\dots,S$.}
        \State{Set $\bm a_t = \hat{\bm\gamma}_t$, where  $\hat{\bm\gamma}_t$ is the OLS coefficient estimate of the linear regression $$\tilde{{y}}_{s,t} =\gamma_{0,t}+ \bm{\gamma}_t^\top\bm{T}(\mathbf{x}_t^{[s]})+\nu_{s,t},$$
        where $S$ is the total number of observations used for estimation.}
        \EndFor
    \end{algorithmic}
    \caption{Calibration of $\bm a$}
    \label{alg:vi2}
\end{algorithm}
Note that Algorithm \ref{alg:vi2} is required in step 6 in Algorithm \ref{alg:vi}. When $j=1$ in Algorithm \ref{alg:vi}, we initialise $\bm{a} =\bm 0$. For $j>1$, we initialise $\bm{a}$ using its latest value. We choose the number of paths $S$ to be equal to three times the dimension of the vector $\bm{a}_t$.

\cite{richard2007efficient} suggest running Algorithm \ref{alg:vi2} iteratively until convergence of $\bm a$ is achieved. In our examples, we find that only one iteration of Algorithm 2 is necessary, and more iterations have no impact on the accuracy of the approach.

\section{Implementation details for the stochastic volatility model}\label{A:sv}
\subsection{Choices of prior}
The prior in the transformed parameter space is given as $p\left(\bm{\theta}\right) = p(\bar{x})p(\kappa)p(c)$, with
\begin{align*}
\text{(i)}\hspace{0.3cm} &p(\bar{x}) = \phi_1\left(\bar{x};0,1000\right),\hspace{1cm}\text{(ii)}\ p(\kappa) = \frac{\exp(\kappa)}{(1+\exp(\kappa))^2},\hspace{1cm}\text{(iii)}\ p(c)\propto e^{-\alpha c}\exp\left(-\frac{\beta}{e^{c}}\right).
\end{align*}
Here, $p(c)$ was constructed by considering an inverse gamma prior on $\sigma^2$, and deriving the corresponding priors on $c$. The shape and rate parameters of the inverse prior are set as $\alpha = 1.001$ and $\beta = 1.001$.
The prior $p(\kappa)$ was constructed by considering a uniform prior on $\rho$  and deriving the corresponding priors on $\kappa$. 

\subsection{Augmented posterior}
The parameters of the SV model are $\bm{\theta}=\left(\bar{x},\kappa,c\right)^\top$. The augmented posterior can be written as
\begin{align*}
p(\bm{\theta},\mathbf{x}|\mathbf{y})\propto&\frac{1}{e^{\frac{x_1}{2}} s}\phi_1\left(\frac{y_1}{e^{\frac{x_1}{2}}}\right)\phi_1\left(\frac{x_1-\bar{x}}{s}\right)\prod_{t=2}^{T}\frac{1}{e^{\frac{x_t}{2}}\sigma}\phi_1\left(\frac{y_t}{e^{\frac{x_t}{2}}}\right)\phi_1\left(\frac{x_t-\bar{x}-\rho(x_{t-1}-\bar{x})}{\sigma}\right)p(\bm{\theta})\,,
\end{align*}
where $s^2 = \frac{\sigma^2}{1-\rho^2}$. Denote $g(\bm\theta,\mathbf{x}) = p(\mathbf{y}|\mathbf{x})p(\mathbf{x}|\bm{\theta})p(\bm{\theta})$. The closed-form expression for $\log g(\bm\theta,\mathbf{x})$ is:
\begin{align*}
\log g(\bm\theta,\mathbf{x})= &\log p(\bm{\theta})-\frac{x_1}{2}-\log(s)-\frac{1}{2}\left(\frac{y_1}{e^{\frac{x_1}{2}}}\right)^2-\frac{1}{2}\left(\frac{x_1-\bar{x}}{s}\right)^2+\\ &\sum_{t=2}^{T}\left[-\frac{x_t}{2}-\log(\sigma)-\frac{1}{2}\left(\frac{y_t}{e^{\frac{x_t}{2}}}\right)^2-\frac{1}{2}\left(\frac{x_t-\bar{x}-\rho(x_{t-1}-\bar{x})}{\sigma}\right)^2\right].
\end{align*}

\subsection{MCMC estimation}
For exact Bayesian inference we implement the following MCMC sampling scheme:\\

\noindent$\underline{\text{Sampling Scheme}}$\\
\ \ \hspace{2cm} Step 1: Generate from $\mathbf{x}|\bm{\theta},\mathbf{y}$.\\ 
\ \ \hspace{2cm} Step 2: Generate from $\bar{x}|\mathbf{x},\mathbf{y},\{\bm{\theta}\backslash\bar{x}\}$.\\
\ \ \hspace{2cm} Step 3: Generate from $\sigma^2|\mathbf{x},\mathbf{y},\{\bm{\theta}\backslash\sigma^2\}$.\\
\ \ \hspace{2cm} Step 4: Generate from $\rho|\mathbf{x},\mathbf{y},\{\bm{\theta}\backslash\rho\}$.\\    
\ \\
For Step 1, we proceed as in \cite{kim1998stochastic}, using a
mixture of seven normals to approximate the distribution of $\log\left[y_t^2e^{-2x_t}\right]$, and then the precision sampler in \cite{chan2009efficient} to generate $\mathbf{x}$.
In Step 2 we use the Gaussian distribution:
 $p(\bar{x}|\mathbf{x},\mathbf{y},\{\bm{\theta}\backslash\bar{x}\}) = \text{N}\left(\mu_{\bar{x}},s_{\bar{x}}^2\right)$
with $s_{\bar{x}}^2 = \left[\frac{1}{1000}+\frac{(T-1)(1-\rho)^2+(1-\rho^2)}{\sigma^2}\right]^{-1}$ and $\mu_{\bar{x}}=s_{\bar{x}}^2\left[\frac{(1-\rho^2)x_1}{\sigma^2}+\frac{(1-\rho)}{\sigma^2}\sum_{t=2}^{T}(x_t-\rho x_{t-1})\right]$. For Step 3 we use the inverse gamma distribution:
$$p(\sigma^2|\mathbf{x},\mathbf{y},\{\bm{\theta}\backslash\sigma^2\}) = \text{IG}\left(\alpha+\frac{T}{2},\beta+\frac{1}{2}\left[(x_1-\bar{x})^2(1-\rho^2)+\sum_{t=2}^T(x_t-\rho x_{t-1}-\bar{x}(1-\rho))^2\right]\right).$$ In Step 4 we use a Metropolis Hastings step, with corresponding proposal $p(\rho) = \text{N}\left(\mu_{\rho},s_{\rho}^2\right) $
where $s_{\rho}^2 = \sigma^2\left[\sum_{t=1}^{T-1}\left(x_t-\bar{x}\right)^2\right]^{-1}$ and $\mu_{\rho} = s_{\rho}^2\frac{\sum_{t=2}^{T}\left(x_t-\bar{x}\right)\left(x_{t-1}-\bar{x}\right)}{\sigma^2}$. Note here that Step 1 can also be employed to generate $p(\mathbf{x}|\mathbf{y},\bm{\theta})$ needed for the hybrid variational Bayes method.

\subsection{Gaussian variational approximation}\label{Sect:Gaussian}
In this section we denote the augmented parameter space of the SV model as $\bm{\psi} = (\bm{\theta}^\top,\mathbf{x}^\top)^\top$. The Gaussian variational approximation considers $$q_\lambda(\bm{\psi}) = q_{\lambda_1}(\bm{\theta})q_{\lambda_2}(\mathbf{x}),$$
with $q_{\lambda_1}(\bm{\theta}) = \phi_d(\bm{\theta};\bm{\mu},BB^\top+D^2)$, $q_{\lambda_2}(\mathbf{x}) = \phi_T(\mathbf{x};\bm{\mu}_x,C_xC_x^\top)$, $C_x$ is a lower triangular Cholesky factor with three bands and $B$ is of dimension $d\times 1$. The variational parameter vectors are $\bm{\lambda}_1 = (\bm{\mu}^\top,\text{vech}(B)^\top,\bm{d}^\top)^\top$, and $\bm{\lambda}_2 = (\bm{\mu}_x^\top,\bm{c}_x^\top)^\top$, where $\bm{d}$ denotes the diagonal elements in $D$, and $\bm{c}_x$ denotes the vector of non-zero elements in $C_x$. 
The ELBO for this approximation is given as 
\begin{align}
    \mathcal{L}(\bm{\lambda}) = E_{\lambda}\left[\log p(\mathbf{y}|\mathbf{x})p(\mathbf{x}|\bm{\theta})p(\bm{\theta})-\log q_\lambda(\bm{\psi}) \right].
\end{align}
The reparametrization gradient of this expression can be computed by writing
\begin{align*}
   \bm{\theta} & = \bm{\mu}+B z+D\bm{\varepsilon}_\theta,\\
    \mathbf{x} & = \bm{\mu}_x+C_x\bm{\varepsilon}_x,
\end{align*}
where $z\sim N(0,1)$ and $\bm{\varepsilon} = (\bm{\varepsilon}_\theta^\top,\bm{\varepsilon}_x^\top)^\top\sim N(\bm{0}_{d+T},I_{d+T})$. Then we get that
\begin{align}
    \nabla_\lambda\mathcal{L}(\bm{\lambda}) = E_{z,\varepsilon}\left[\frac{\partial \bm{\psi}}{\partial\bm{\lambda}}^\top\left[\nabla_\psi\log p(\mathbf{y}|\mathbf{x})p(\mathbf{x}|\bm{\theta})p(\bm{\theta})-\nabla_\psi\log q_\lambda(\bm{\psi})\right] \right].
\end{align}
Note that $\frac{\partial \bm{\psi}}{\partial\bm{\lambda}} = \text{blockdiag}(\frac{\partial \bm{\theta}}{\partial\bm{\lambda}_1},\frac{\partial \mathbf{x}}{\partial\bm{\lambda}_2})$, where the operator blockdiag indicates the diagonal stacking of two matrices, $\frac{\partial \bm{\theta}}{\partial\bm{\lambda}_1}$ was provided in a previous section, and $\frac{\partial \mathbf{x}}{\partial\bm{\lambda}_2} = \left[I_T \hspace{0.2cm} (\bm{\varepsilon}_x^\top\otimes I_T)P\right]$, where $P$ is a matrix such that $\frac{\partial \mathbf{x}}{\partial\bm{c}_x} = \frac{\partial \mathbf{x}}{\partial C_x}P$. Additionally, note that $\nabla_\psi\log q_\lambda(\bm{\psi}) = (\nabla_\theta\log q_{\lambda_1}(\bm{\theta})^\top,\nabla_x\log q_{\lambda_2}(\mathbf{x})^\top)^\top$. An expression for $\nabla_\theta\log q_{\lambda_1}(\bm{\theta})^\top$ is provided in \cite{ong2018gaussian}, while $\nabla_x\log q_{\lambda_2}(\mathbf{x}) = -(C_xC_x^\top)^{-1}(\mathbf{x}-\bm{\mu}_x)$. The Gaussian variational approximation is calibrated by using an unbiased estimate of this ELBO gradient inside an SGA algorithm. 

\subsection{Required gradients}
The VB methods require the gradient $\nabla_{\theta}\log g(\bm\theta,\mathbf{x}) = \nabla_{\theta}\log p(\mathbf{y}|\mathbf{x})p(\mathbf{x}|\bm{\theta})p(\bm{\theta})$. Note that the derivatives of the priors with respect to their corresponding arguments are: 
\begin{align*}
\text{(i)}\hspace{0.3cm} &\frac{\partial \log p(\bar{x})}{\partial\bar{x}}=-\frac{\bar{x}}{1000},\hspace{1cm}\text{(ii)}\ \frac{\partial \log p(\kappa)}{\partial\kappa}=1-2\frac{\rho}{0.995},\hspace{1cm}\text{(iii)}\ \frac{\partial \log p(c)}{\partial c}=-\alpha+\beta e^{-c}.
\end{align*}
With these derivatives we can then construct $$\nabla_{\theta}\log g(\bm\theta,\mathbf{x}) = \left(\nabla_{\bar{x}}\log g(\bm\theta,\mathbf{x}),\nabla_{\kappa}\log g(\bm\theta,\mathbf{x}),\nabla_{c}\log g(\bm\theta,\mathbf{x})\right)^\top,$$
with each of its elements defined as:
\begin{align*}
\nabla_{\bar{x}}\log g(\bm\theta,\mathbf{x}) &=\frac{\partial \log p(\bar{x})}{\partial\bar{x}}+\frac{x_1-\bar{x}}{s^2}-\sum_{t=2}^{T}\frac{(\rho-1)}{\sigma}\left[\frac{x_t-\bar{x}-\rho(x_{t-1}-\bar{x})}{\sigma}\right],\\
\nabla_{\kappa}\log g(\bm\theta,\mathbf{x}) &=\frac{\partial \log p(\kappa)}{\partial\kappa}+\left\{\frac{\rho}{1-\rho^2}\left[\frac{(x_1-\bar{x})^2}{s^2}-1\right]+\sum_{t=2}^{T}\frac{x_{t-1}-\bar{x}}{\sigma^2}\left[x_t-\bar{x}-\rho(x_{t-1}-\bar{x})\right]\right\}\frac{0.995\exp(\kappa)}{(1+\exp(\kappa))^2},\\
\nabla_{c}\log g(\bm\theta,\mathbf{x}) &=\frac{\partial \log p(c)}{\partial c}-\frac{1}{2}+\frac{(x_1-\bar{x})^2}{2s^2}+\sum_{t=2}^{T}-\frac{e^{\frac{c}{2}}}{2\sigma}+\frac{\left[x_t-\bar{x}-\rho(x_{t-1}-\bar{x})\right]^2e^{\frac{c}{2}}}{2\sigma^3}.
\end{align*}

\section{Additional results from the numerical experiments}\label{A:numerical}
\begin{figure}[H]
\caption{Posterior intervals using different update frequencies of $q(\mathbf{x}|\mathbf{y})$}
\centering
\includegraphics*[width=.9\textwidth]{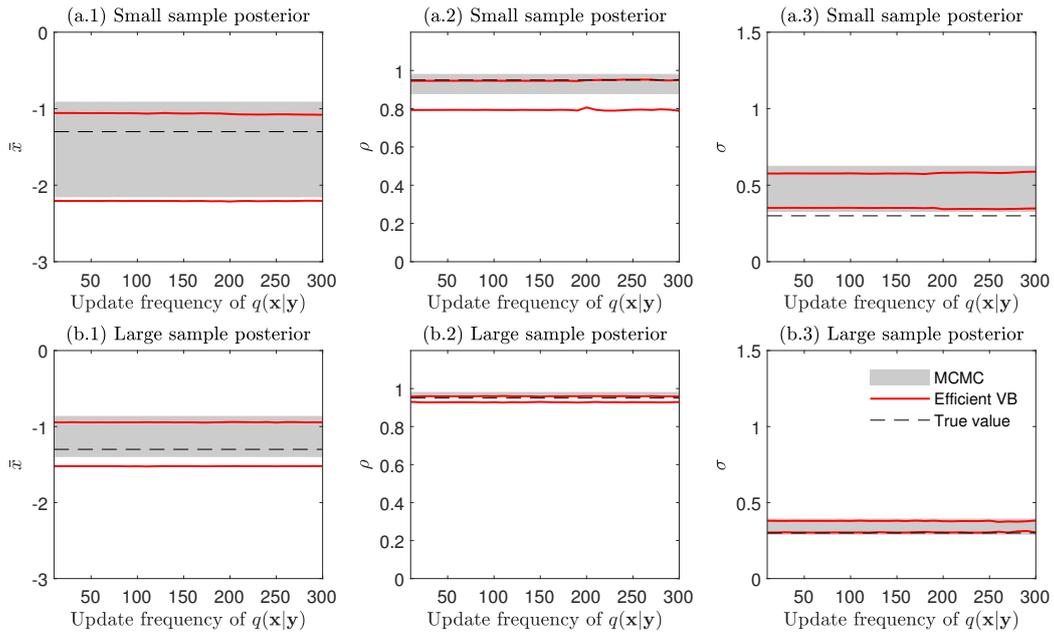}\\[3mm]
\begin{flushleft}
This figure shows the 99\% posterior intervals for the parameters of the model. The first row corresponds to a sample size of $T = 500$. The second row corresponds to a sample size of $T=4000$. The gray areas indicate the posterior intervals for MCMC, while the red lines indicate those for Efficient VB. The horizontal dashed lines indicate the true values in the data generating process.
\end{flushleft}
\label{fig:SVlargeApp}
\end{figure}

\section{Implementation details for the Skellam application}\label{A:skellam}
To conduct variational Bayes the parameters have to be transformed into the real line. Denote $L^*$ to be a matrix with identical off-diagonal values to the lower triangular matrix $L$ but whose positive diagonal values are $L_{i,i}^* = log(L_{i,i})$. Transform $\kappa_i$ into the real line via the logistic transformation so that $\kappa_i = \text{logit}^{-1}(\varkappa_i)$. Transform $\omega_i$ into the real line via the logistic transformation so that ${\omega}_i = \text{logit}^{-1}(\tilde{\omega}_i)$. Denote $\bm{\varkappa} = (\varkappa_1,\dots,\varkappa_N)^\top$, $\tilde{\bm{\omega}} = (\tilde{\omega}_1,\dots,\tilde{\omega}_N)^\top$ and $\bm{l} = \text{vech}(L^*)$ and the complete parameter vector of the model can be written as $\bm{\theta} = (\bm{\varkappa}^\top,\bar{\bm x}^\top,\bm{l}^\top,\tilde{\bm{\omega}}^\top,\bm{\beta}_1^\top,\dots,\bm{\beta}_N^\top)^\top$. The priors $p(\varkappa_i) = \frac{\exp(\varkappa_i)}{(1+\exp(\varkappa_i))^2}$, $p(\tilde{\omega}_i) = \frac{\exp(\tilde{\omega}_i)}{(1+\exp(\tilde{\omega}_i))^2}$ and $p(L^*) = |LL^\top|^{-\frac{d+1}{2}}\prod_{i=1}^d\exp(L^*_{i,i})(d+2-i)$ imply respectively uniform priors on $\kappa_i$ and $\tilde{\omega}_i$ in the unit interval, and a Jeffrey's prior in the covariance matrices $(LL^\top)^{-1}$.







\subsection{Required gradients for Efficient VB}\label{sec:gradapp}
Below we denote
$$\log p(\xvec_t|\xvec_{t-1},\bm{\theta}) =-\frac{1}{2}d\log(2\pi)+\frac{1}{2}\log(|LL^\top|) -\frac{1}{2}(\xvec_t-\bm{\eta}_t)^\top LL^\top(\xvec_t-\bm{\eta}_t),$$
where $\bm{\eta}_t = \bar{\bm x}+\Omega\xvec_{t-1}$ and $\xvec_0 = (\log\text{VAR}(y_1),\dots,\log \text{VAR}(y_N)))^\top$ .

Efficient VB evaluates the gradient $\nabla_\theta\log p(\yvec,\xvec|\bm\theta)p(\bm\theta) = \nabla_\theta\log p(\yvec,\xvec|\bm\theta)+\nabla_\theta\log p(\bm\theta)$, where 
{\scriptsize
\begin{align*}
\nabla_\theta\log p(\yvec,\xvec|\bm\theta) = &\left(\frac{\partial }{\partial \bm\varkappa} \log p(\yvec|\xvec,\bm{\theta}),\frac{\partial }{\partial \bm{\bar{x}}} \log p(\xvec|\bm{\theta}),\frac{\partial }{\partial \bm{l}} \log p(\xvec|\bm{\theta}),\frac{\partial }{\partial \tilde{\bm{\omega}}} \log p(\xvec|\bm{\theta}),\frac{\partial }{\partial \bm{\beta}_1} \log p(\yvec|\xvec,\bm{\theta}),\dots,\frac{\partial }{\partial \bm{\beta}_N} \log p(\yvec|\xvec,\bm{\theta})\right)^\top\\
\nabla_\theta\log p(\bm\theta) &= \left(\frac{\partial}{\partial \bm\varkappa}\log p(\bm\varkappa),\frac{\partial}{\partial \bar{\xvec}}\log p(\bar{\xvec}),\frac{\partial}{\partial \bm{l}}\log p(L^*),\frac{\partial}{\partial \tilde{\bm\omega}}\log p(\tilde{\bm\omega}),\frac{\partial}{\partial \bm{\beta}_1}\log p( \bm{\beta}_1),\dots,\frac{\partial}{\partial \bm{\beta}_N}\log p( \bm{\beta}_N)\right)^\top.
\end{align*}
}
The elements of $\nabla_\theta\log p(\yvec,\xvec|\bm\theta)$ are computed as
\begin{align*}
i) & \ \ \ \frac{\partial }{\partial \bm\varkappa} \log p(\yvec|\xvec,\bm{\theta}) = \sum_{t=1}^T \frac{\partial }{\partial \bm\varkappa} \log p(\yvec_t|\xvec_t,\bm{\theta});  &ii) \ \ \ \ \frac{\partial }{\partial \bm{\bar{x}}} \log p(\xvec|\bm{\theta})&=\sum_{t=1}^T\frac{\partial }{\partial \bar{\bm{x}}}\log p(\xvec_t|\xvec_{t-1},\bm{\theta});\\
iii)& \ \ \ \frac{\partial }{\partial \bm{l}} \log p(\xvec|\bm{\theta})=\sum_{t=1}^T\frac{\partial }{\partial \bm{l}}\log p(\xvec_t|\xvec_{t-1},\bm{\theta});  &iv) \ \ \ \
\frac{\partial }{\partial \tilde{\bm{\omega}}} \log p(\xvec|\bm{\theta})&=\sum_{t=1}^T\frac{\partial }{\partial \tilde{\bm{\omega}}}\log p(\xvec_t|\xvec_{t-1},\bm{\theta});\\
v) & \ \ \ \frac{\partial }{\partial \bm{\beta}_i} \log p(\yvec|\xvec,\bm{\theta}) = \sum_{t=1}^T \frac{\partial }{\partial \bm{\beta}_i} \log p(\yvec_t|\xvec_t,\bm{\theta}),
\end{align*}
with
{\scriptsize
\begin{align*}
i) & \ \ \ \ \frac{\partial }{\partial \bm\varkappa} \log p(\yvec_t|\xvec_t,\bm{\theta})= \left(\frac{\delta_{0}(y_{1,t})-\exp\left(-\sigma^2_{1,t}\right)\mathcal{I}_{|y_{1,t}|}(\sigma^2_{1,t})}{p(y_{1,t}|x_{1,t},\bm{\theta})},\dots,\frac{\delta_{0}(y_{N,t})-\exp\left(-\sigma^2_{N,t}\right)\mathcal{I}_{|y_{N,t}|}(\sigma^2_{N,t})}{p(y_{N,t}|x_{N,t},\bm{\theta})}\right)^\top\frac{\partial \bm{\kappa}}{\partial \bm{\varkappa}};\\
ii) & \ \ \ \ \frac{\partial }{\partial \bar{\bm{x}}}\log p(\xvec_t|\xvec_{t-1},\bm{\theta}) = (\xvec_t-\bm{\eta}_t)^\top LL^\top;\\
iii) & \ \ \ \ \frac{\partial }{\partial \bm{l}} \log p(\xvec_t|\xvec_{t-1},\bm{\theta})= \frac{1}{2}\text{vec}((LL^\top)^{-1})^\top(I_{d^2}+K_{d,d})(L\otimes I_d)R-\frac{1}{2}((\xvec_t-\bm{\eta}_t)^\top\otimes(\xvec_t-\bm{\eta}_t)^\top)(I_{d^2}+K_{d,d})(L\otimes I_d)R;\\
iv) & \ \ \ \ \frac{\partial }{\partial \tilde{\bm{\omega}}}\log p(\xvec_t|\xvec_{t-1},\bm{\theta}) = (\xvec_t-\bm{\eta}_t)^\top LL^\top\text{diag}(\xvec_{t-1})\frac{\partial \bm{\omega}}{\partial \tilde{\bm{\omega}}};\\
v) & \ \ \ \ \frac{\partial }{\partial \bm\beta_i} \log p(\yvec_t|\xvec_t,\bm{\theta})= {\frac{(1-\kappa_i)(\exp\left(-\sigma^2_{i,t}\right)\mathcal{I}_{|y_{i,t}|}'(\sigma^2_{i,t})-\mathcal{I}_{|y_{i,t}|}(\sigma^2_{i,t})\exp\left(-\sigma^2_{i,t}\right))\sigma^2_{i,t}\bm{w}_t^\top}{p(y_{i,t}|x_{i,t},\bm{\theta})}}.
\end{align*}
}
The elements of $\nabla_\theta\log p(\bm\theta)$ are computed as

\begin{align*}
i) & \ \ \ \ \frac{\partial}{\partial \bm\varkappa}\log p(\bm\varkappa) = (\bm{1}_N-2\bm{\kappa})^\top; & ii) \ \ \ \ \frac{\partial}{\partial \bar{\xvec}}\log p(\bar{\xvec}) = -\frac{1}{100}\bar{\xvec};
\end{align*}
\begin{align*}
iii) &\ \ \ \ \frac{\partial}{\partial \bm{l}}\log p(L^*) = -\frac{d+1}{2}  \text{vec}((LL^\top)^{-1})^\top(I_{d^2}+K_{d,d}) (L\otimes I_d)R+\bm{\vartheta}^\top;
\end{align*}
\begin{align*}
iv) & \frac{\partial}{\partial \tilde{\bm{\omega}}}\log p(\tilde{\bm{\omega}}) = (\bm{1}_N-2\bm{\omega})^\top; & v) \ \ \ \ \frac{\partial}{\partial \bm{\beta}_i}\log p(\bm{\beta}_i) = -\frac{1}{100}\bar{\bm{\beta}_i}.
\end{align*}
Here, $\bm{\vartheta}$ is a column vector with $N^2$ elements, so that $\bm{\vartheta}_{((s-1)(N+1)+1)}= N+2-s$ for $s\in \{1,\dots,N\}$. The matrix $R$ is a $N^2\times N^2$ diagonal matrix with elements  $R_{((s-1)(N+1)+1,(s-1)(N+1)+1)}=e^{L^*_{s,s} }$ for $s\in \{1,\dots,N\}$, and ones elsewhere.

The derivative $\frac{\partial }{\partial \bm\beta_i} \log p(\yvec_t|\xvec_t,\bm{\theta})$ can be further simplified as follows. Whenever $y_{i,t}\ne 0$,
\begin{align}
\frac{\partial }{\partial \bm\beta_i} \log p(\yvec_t|\xvec_t,\bm{\theta}) &= {\frac{(1-\kappa_i)(\widetilde{\mathcal{I}}_{|y_{i,t}|-1}(\sigma^2_{i,t})-\frac{|y_{i,t}|}{\sigma^2_{i,t}}\widetilde{\mathcal{I}}_{|y_{i,t}|}(\sigma^2_{i,t}) 
   -\widetilde{\mathcal{I}}_{|y_{i,t}|}(\sigma^2_{i,t}))\sigma^2_{i,t}\bm{w}_t^\top}{p(y_{i,t}|x_{i,t},\bm{\theta})}},
\end{align}
and when $y_{i,t}= 0$,
\begin{align}
\frac{\partial }{\partial \bm\beta_i} \log p(\yvec_t|\xvec_t,\bm{\theta})&= {\frac{(1-\kappa_i)(\widetilde{\mathcal{I}}_{1}(\sigma^2_{i,t})
   -\widetilde{\mathcal{I}}_{|y_{i,t}|}(\sigma^2_{i,t}))\sigma^2_{i,t}\bm{w}_t^\top}{p(y_{i,t}|x_{i,t},\bm{\theta})}},
\end{align}
with $\widetilde{\mathcal{I}}_{\nu}(z) = {\mathcal{I}}_{\nu}(z)\exp(-|z|))$.

\subsection{PMCMC}
To implement PMCMC we evaluate an approximate estimate of the likelihood function using a bootstrap particle filter with a total of 1000 particles. We then use this estimate inside a Metropolis-Hastings scheme. At the beginning of each step, the parameter space is split into two blocks $\bm{\theta}_1$ and $\bm{\theta}_2$, with the elements of the two blocks selected at random. At the ith step, we set  $\bm{\theta}_1^{(i)} = \bm{\theta}_1^{(\text{new})}$ with probability 
$$\alpha = \text{min}\left(\frac{\hat{p}(\yvec|\bm{\theta}_1^{(\text{new})},\bm{\theta}_2^{(i-1)})p(\bm{\theta}_1^{(\text{new})})}{\hat{p}(\yvec|\bm{\theta}_1^{(i-1)},\bm{\theta}_2^{(i-1)})p(\bm{\theta}_1^{(i-1)})},0\right),$$
where $\bm{\theta}_1^{(\text{new})}$ is a candidate from the Gaussian proposal distribution $\bm{\theta}_1|\bm{\theta}_1^{(i-1)} \sim N(\bm{\theta}_1^{(i-1)},\text{diag}(\bm{steps}_1))$. Here, $\bm{steps}_1$ is a vector of step sizes for $\bm{\theta}_1$ set adaptively to achieve acceptance rates between 15\% and 30\%. If the candidate draw is rejected, we set 
$\bm{\theta}_1^{(i)} = \bm{\theta}_1^{(i-1)}$. A similar Metropolis-Hastings step is then undertaken for the second block of parameters.

\subsection{Gaussian approximation}
Since we only implement Gaussian VB for the univariate Skellam model, we remove the variable index $i$ in the notation below. The Gaussian approximation is identical to the one in Section~\ref{Sect:Gaussian}.
Implementation of the approach requires the gradient 
$$\nabla_\psi\log p(\mathbf{y}|\mathbf{x},\bm{\theta})p(\mathbf{x}|\bm{\theta})p(\bm{\theta})=([\nabla_\theta\log p(\mathbf{y}|\mathbf{x},\bm{\theta})p(\mathbf{x}|\bm{\theta})p(\bm{\theta})]^\top,[\nabla_\xvec\log p(\mathbf{y}|\mathbf{x},\bm{\theta})p(\mathbf{x}|\bm{\theta})]^\top)^\top.$$
The gradient $\nabla_\theta\log p(\mathbf{y}|\mathbf{x},\bm{\theta})p(\mathbf{x}|\bm{\theta})p(\bm{\theta})$ is provided in Section~\ref{sec:gradapp}. The gradient $\nabla_\xvec\log p(\mathbf{y}|\mathbf{x},\bm{\theta})$ is a $T-$dimensional vector, where its $t^{\text{th}}$ element is $${\frac{(1-\kappa)(\exp\left(-\sigma^2_{t}\right)\mathcal{I}_{|y_{t}|}'(\sigma^2_{t})-\mathcal{I}_{|y_{t}|}(\sigma^2_{t})\exp\left(-\sigma^2_{t}\right))\sigma^2_{t}}{p(y_{t}|x_{t},\bm{\theta})}}.$$ 
The gradient $\nabla_\xvec\log p(\mathbf{x}|\bm{\theta})$ is a $T-$dimensional vector, where its element $t\in{\{1,\dots,T-1\}}$ is 
$$-(x_t-\eta_t)(L^2)+\omega(L^2)(x_{t+1}-\eta_{t+1}),$$
and its $T^{\text{th}}$ element is $-(x_T-\eta_T)(L^2).$

\section{Additional results from the Skellam empirical application}\label{A:skellam_results}
\begin{figure}[H]
\caption{ELBO for the univariate Skellam stochastic volatility models}
\centering
\includegraphics*[width=\textwidth]{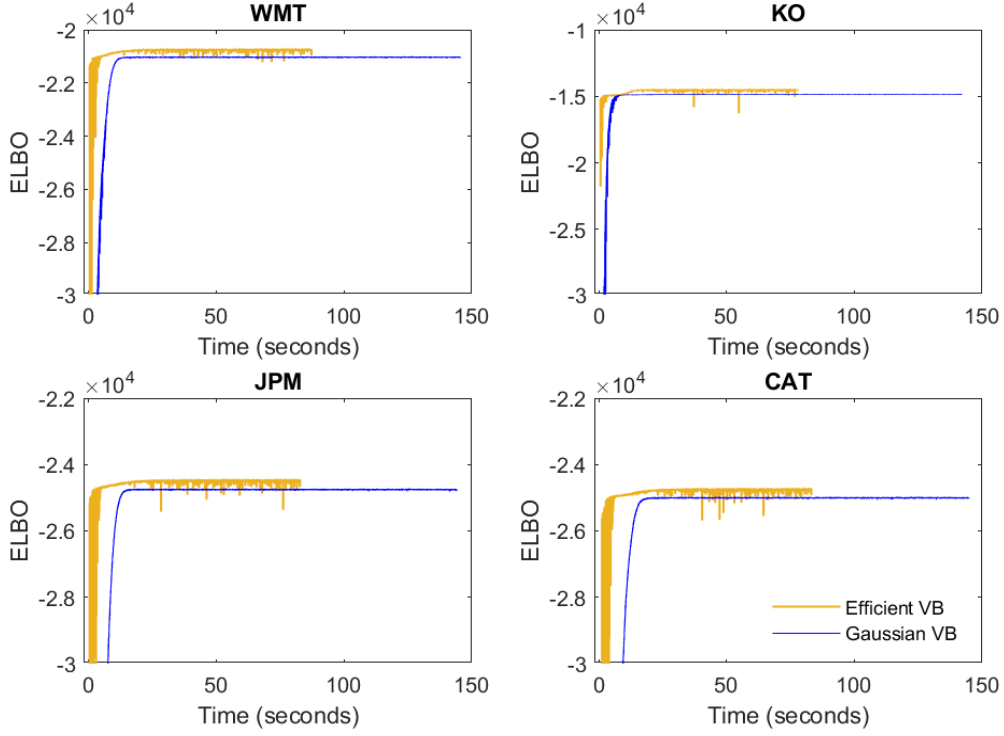}
\begin{flushleft}
This figure presents the ELBO traces for each stock with the univariate Skellam stochastic volatility models. The yellow lines correspond to Efficient VB while the blue lines correspond to Gaussian VB.
\end{flushleft}
\label{fig:SkellamELBO}
\end{figure}


\section{Implementation details for the TVP-VAR-SV model}\label{A:var}
This section demonstrates how the TVP-VAR-SV model in \eqref{eq:tvpvarsv} can be represented by the $N$ unrelated equations in \eqref{eq:ssm_tvpvarsv}. First, pre-multiply~\eqref{eq:tvpvarsv} 
by $L_t$, so that 
\begin{eqnarray*}
	L_t\yvec_t &= &L_t\betavec_{0,t}+\sum_{s=1}^p L_tB_{s,t}\yvec_{t-s}+\epsilonvec_t 
	= \gammavec_{0,t}+\sum_{s=1}^p\Gamma_{s,t}\yvec_{t,s}+\epsilonvec_t,
\end{eqnarray*}
where $\Gamma_{s,t} = L_tB_{s,t}$ and $\gamma_{0,t} = L_t B_{0,t}$.
Denote the 
non-fixed elements of the $i$th row of $L_t$ as $\bm{l}_{1:i-1,t} = \left(l_{i,1,t},\dots,l_{i,i-1,t}\right)^\top$
for $i\geq 2$, so that the entire $i$th row of $L_t$ is $(\bm{l}_{1:i-1,t}^\top,1,\bm{0}_{N-i}^\top)$.
Then, each of the $i=1,\ldots,N$ individual equations of the model can be written as 
\begin{equation}
	y_{i,t}+\mathbf{y}_{1:i-1,t}^\top\bm{l}_{1:i-1,t} =\left(\mathbf{y}_{t-1}^\top,\dots,\mathbf{y}_{t-p}^\top,1\right)\bm{\gamma}_{i,t}+\epsilon_{i,t},
	\label{eq:tvpvarsveqi}
\end{equation}
where  $\mathbf{y}_{1:i-1,t} = \left(y_{1,t},\dots,y_{i-1,t}\right)^\top$, $\bm{\gamma}_{i,t} = \left(\Gamma_{i,1,t},\dots,\Gamma_{i,p,t},\gamma_{i,0,t}\right)^\top$,
$\Gamma_{i,s,t}$ denotes the $i$th row of  $\Gamma_{s,t}$,
$\gamma_{i,0,t}$ is the $i$th element in $\bm{\gamma}_{0,t}$, and $\epsilon_{i,t}\sim N(0,\exp(h_{i,t}))$. 
The $i$th equation can be simplified to
\[
y_{i,t}=\zvec_{i,t}^\top\etavec_{i,t}+\epsilon_{i,t},
\]
where $\zvec_{i,t}^\top=\left(\mathbf{y}_{t-1}^\top,\dots,\mathbf{y}_{t-p}^\top,1,-\mathbf{y}_{1:i-1,t}^\top\right)$
and $\etavec_{i,t}^\top = \left(\gammavec_{i,t}^\top,\bm{l}_{1:i-1,t}^\top\right)$. The $\etavec_{i,t}$ state vector notation is not to be confused with the function $\bm\eta(.,.)$ employed to define exponential density functions in Section \ref{sec:vb}.
In this representation, the coefficient vector is of dimension $Np+i=J_i/2$ and follows the 
random walk
$\etavec_{i,t}=\etavec_{i,t-1}+\diag({\bm{\alpha}_{2,i}})\tilde{\varepsilonvec}_{i,t}$,
with $\tilde{\varepsilonvec}_{i,t}\sim N(\bm{0},I_{J_i/2})$. 

The coefficients $\etavec_{i,t}$ are  
transformed to the ``non-centered'' representation 
$\etavec_{i,t}=\bm{\alpha}_{1,i}+\mbox{diag}(\bm{\alpha}_{2,i})\widetilde\etavec_{i,t}$ as a sum of a time-invariant term $\bm{\alpha}_{1,i}$
and scaled time-varying deviations $\widetilde\etavec_{i,t}$. 
Substituting in this parameterization gives
the state space model
\begin{eqnarray}
	y_{i,t} &= & ({\mathbf{z}}_{i,t}^\top,{\mathbf{z}}_{i,t}^\top\text{diag}(\tilde{\bm{\eta}}_{i,t}))\bm{\alpha}_i+\epsilon_{i,t}, \nonumber\\
 \tilde{\bm{\eta}}_{i,t}&=&\tilde{\bm{\eta}}_{i,t-1}+\tilde{\varepsilonvec}_{i,t},\nonumber\\
 h_{i,t}& = & \bar{h}_i+\rho(h_{i,t-1}-\bar{h}_i)+e_{i,t},\mbox{ for }
i=1,\ldots,N,
 \label{eq:apptvp1}
\end{eqnarray}
with $\alphavec_i^\top=(\bm{\alpha}_{1,i}^\top,\bm{\alpha}_{2,i}^\top)$.
\cite{huber2021inducing} employ a horseshoe prior for the vector of coefficients $\alphavec_i = \left(\alpha_{i,1},\dots,\alpha_{i,J_i}\right)^\top$, which can be represented as
\begin{align}
\alpha_{i,j}|\xi_i\chi_{i,j}~\sim N(0,\xi_i\chi_{i,j}),\qquad  \chi_{i,j}|\nu_{i,j}\sim \mathcal{G}^{-1}\left(\frac{1}{2},\frac{1}{\nu_{i,j}}\right),\\  \xi_i|\kappa_{i}\sim \mathcal{G}^{-1}\left(\frac{1}{2},\frac{1}{\kappa_{i}}\right),\qquad  \nu_{i,1},\dots,\nu_{i,J_i},\kappa_{i}\sim \mathcal{G}^{-1}\left(\frac{1}{2},1\right).
\end{align}

As noted in \cite{ingraham2017variational}, the horseshoe prior can lead to funnel-shaped posteriors for the elements of $\bm{\alpha}_i$, which cannot be approximated using Gaussian distributions. This hinders the performance of Gaussian variational inference. The authors suggest to transform $\bm{\alpha}_i$ into $\bm{\tau}_i = \left(\tau_{i,1},\dots,\tau_{i,J_i}\right)^\top$, where $\alpha_{i,j} = \tau_{i,j}\sqrt{\xi_i\chi_{i,j}}$ with prior $\tau_{i,j}\sim N(0,1)$. Unlike $\bm{\alpha}_i$, the posterior distributions for the elements of $\bm{\tau}_i$ can be easily approximated by Gaussians. Note that $\bm{\alpha}_i = \sqrt{\xi_i}(\bm{\tau}_i\circ\sqrt{\bm{\chi}_i})$, where `$\circ$' denotes the Hadamard product. We can replace this expression into \eqref{eq:apptvp1} to obtain the measurement equation in \eqref{eq:ssm_tvpvarsv}. The state equation in \eqref{eq:ssm_tvpvarsv} can be recovered by stacking the latent states in the single vector $\xvec_{i,t} = (\tilde{\etavec}_{i,t}^\top,h_{i,t})^\top$, such that 
$$\mathbf{x}_{i,t} = \bar{\mathbf{x}}_i+A_{1,i}\mathbf{x}_{i,t-1}+A_{2,i}\bm{\varepsilon}_{i,t},
$$
where 
$$\bar{\mathbf{x}}_i =\begin{bmatrix}
\bm{0}_{J_i/2}  \\
\bar{h}_i(1-\rho_i)
\end{bmatrix}, \text{ } A_{1,i} = \begin{bmatrix}
I_{J_i/2} & \bm{0}_{J_i/2} \\
\bm{0}_{J_i/2}^\top & \rho_i
\end{bmatrix} \text{, } A_{2,i}= \begin{bmatrix}
I_{J_i/2} & \bm{0}_{J_i/2} \\
\bm{0}_{J_i/2}^\top & \sigma_i
\end{bmatrix},$$
and $\varepsilonvec_{i,t}\sim N(\bm{0},I_{(J_i/2)+1}).$

\section{Additional results from the TVP-VAR-SV application}\label{A:application}
\begin{figure}[h!]
\caption{ELBO for the TVP-VAR-SV model}
\centering
\includegraphics*[scale = 0.63,trim={0cm 1.5cm 0 1.3cm}]{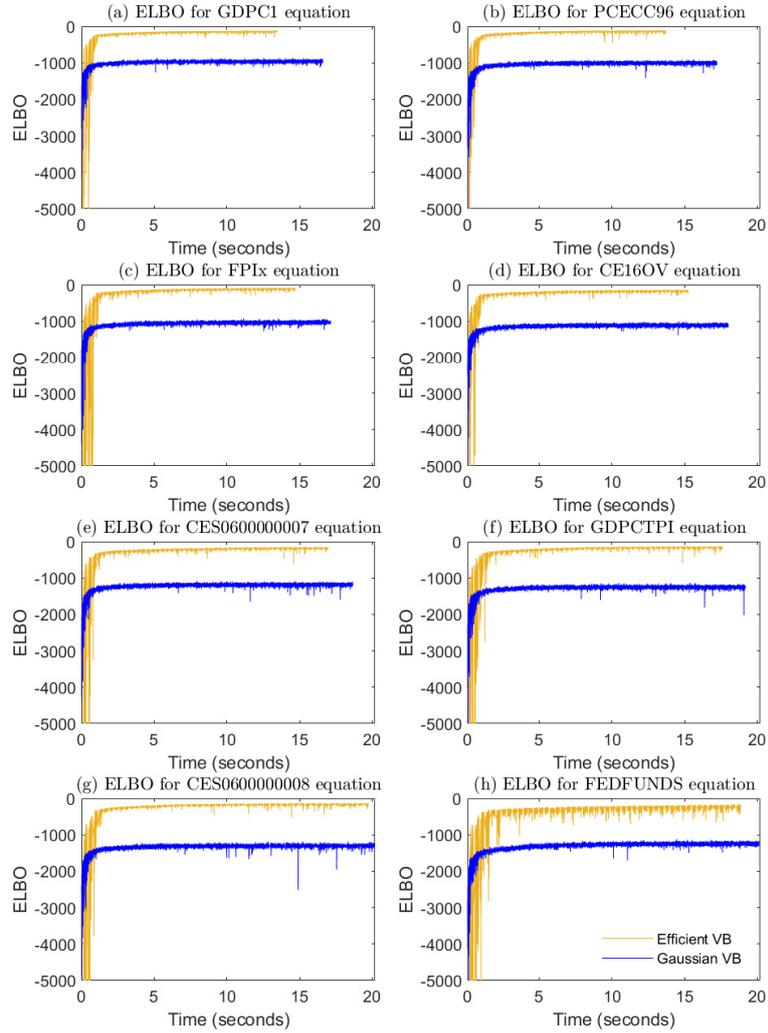}
\begin{flushleft}
This figure presents the ELBO traces for each of the equations in the TVP-VAR-SV. The yellow lines correspond to Efficient VB while the blue lines correspond to Gaussian VB.
\end{flushleft}
\label{fig:VARELBO}
\end{figure}

\begin{figure}[tb!]
\caption{Posterior mean of the time-varying volatilities in the TVP-VAR-SV}
\centering
\includegraphics[scale=0.7]{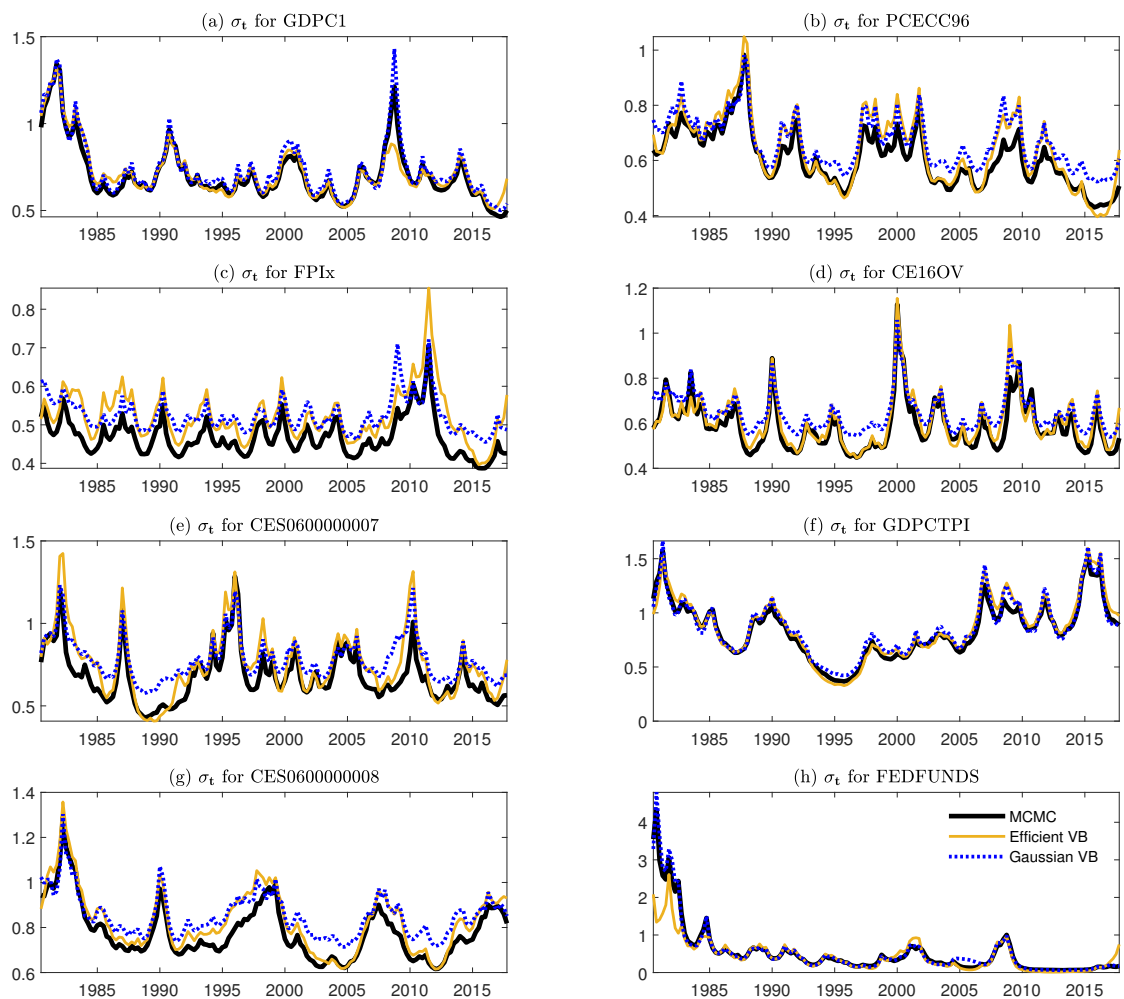}\\[3mm]
\begin{flushleft}
This figure shows the posterior means of $\exp(h_{4,t}/2)$ in \eqref{eq:tvpvarsv} across time $t$, for Efficient VB, Gaussian VB, and MCMC, indicated by the solid yellow, dotted blue, and solid black line, respectively.
\end{flushleft}
\label{fig:VolstatesAll}
\end{figure}

\end{document}